\documentclass[symmetry,article,accept,oneauthors,latex,12pt,a4paper]{mdpi}

\lastpage{25}
\doinum{10.3390/sym11080994}
\pubvolume{11}
\pubyear{2019}
\history{Received: 5 June 2019 / Revised: 25 July 2019 / Accepted: 26 July 2019 / Published: 3 August 2019 }

\usepackage{graphicx}

\newcommand{\myskip}[1]{}

\usepackage[english]{babel}
 \usepackage{amsmath}
\usepackage{amssymb}
 \usepackage{graphicx}
 \usepackage{dcolumn}
\usepackage{bm}
\usepackage{epstopdf}
\usepackage{slashed} 

\newcommand{\refl}{\ref}

\newcommand{\ta}{{\tilde a}}
\newcommand{\ts}{{\tilde s}}

\newcommand{\vA}{{\bf A}}
\newcommand{\ve}{{\bf e} }
\newcommand{\vG}{{\bf G}}
\newcommand{\vL}{{\bf L}}
\newcommand{\vM}{{\bf M}}
\newcommand{\vq}{{\bf q}}
\newcommand{\vQ}{{\bf Q}}

\newcommand{\vY}{{\bf Y}}
\newcommand{\vell}{\bm{\ell}}
\newcommand{\vnu}{\bm{\nu}}

\newcommand{\mb}{{\bar m}}
 
\newcommand{\sqt}{\sqrt{2}}

\newcommand{\ci}{\,\text{\cal i}}

\newcommand{\MM}{{M^M_R}}
\newcommand{\mM}{M^N}

\newcommand{\ub}{{\bar u}}

\newcommand{\mub}{{\bar \mu}}
\newcommand{\mube}{{\bar\mu_1}}
\newcommand{\mubt}{{\bar\mu_2}}
\newcommand{\mubd}{{\bar\mu_3}}
\newcommand{\mubi}{{\bar\mu_i}}
\newcommand{\Deltae}{\bar\Delta_1}
\newcommand{\Deltat}{\bar\Delta_2}
\newcommand{\Deltad}{\bar\Delta_3}
\newcommand{\Deltai}{\bar\Delta_i}
\newcommand{\Deltak}{\bar\Delta_k}

\newcommand{\mbe}{\bar m_1}
\newcommand{\mbt}{\bar m_2}
\newcommand{\mbd}{\bar m_3}
\newcommand{\mbi}{\bar m_i}
\newcommand{\Mbe}{\bar M_1}
\newcommand{\Mbt}{\bar M_2}
\newcommand{\Mbd}{\bar M_3}
\newcommand{\Mbi}{\bar M_i}

\newcommand{\cc}{{\,c}}

\newcommand{\cA}{{\cal A}}
\newcommand{\cS}{{\cal S}}
\newcommand{\cC}{{\cal C}}
\newcommand{\cCdag}{{\cal C}^\dagger}

\newcommand{\cL}{{\cal L}}

\newcommand{\hc}{h.c.}

\newcommand{\cH}{{\cal H}}
\newcommand{\cM}{{\cal M}}

\newcommand{\cV}{{\cal V}}

\newcommand{\diag}{{\rm diag}}

\newcommand{\BEQ}{\begin{eqnarray}}
\newcommand{\EEQ}{\end{eqnarray}}
\newcommand{\BEA}{\begin{eqnarray}}
\newcommand{\EEA}{\end{eqnarray}}
\newcommand{\nn}{\nonumber}

\newcommand{\bn}{{\bf n}}
\newcommand{\bN}{{\bf N}}

\newcommand{\p}{\partial}

\newcommand{\eV}{{\rm eV}}
\newcommand{\MeV}{{\rm MeV}}
\newcommand{\GeV}{{\rm GeV}}

\newcommand{\vp}{{\bf p}}
\newcommand{\vr}{{\bf r}}
\newcommand{\ccdot}{\!\cdot\!}
\newcommand{\mmin}{\!-\!}
\newcommand{\pplus}{\!+\!}

\newcommand{\half}{\frac{1}{2}}

\newcommand{\veen}{{1}}

\newcommand{\mD}{\bar m}
\newcommand{\rmL}{{\rm L}}

\newcommand{\mDe}{\bar m_1}
\newcommand{\mDt}{\bar m_2}
\newcommand{\mDd}{\bar m_3}

\renewcommand{\vM}{M}
\renewcommand{\vY}{Y}
\newcommand{\DpM}{{\DpM}}
\renewcommand{\DpM}{{DM}}

\newcommand{\DMnSM}{\rm DM$\nu$SM}
\newcommand{\SnuM}{\rm S$\nu$M}

\newcommand{\voetnoot}{\footnote}

\Title{The Standard Model of particle physics with Diracian neutrino sector}

\Author{Theodorus Maria Nieuwenhuizen}

\address{Institute for Theoretical Physics, University of Amsterdam, Science Park 904,  1098 XH Amsterdam, The Netherlands} 
\corres{e-mail: t.m.nieuwenhuizen\@ uva.nl;  telephone +31-20-525-6332}

\abstract{
The minimally extended standard model of particle physics contains three right handed or sterile neutrinos, coupled to the active ones by a Dirac mass matrix and mutually by a Majorana mass matrix. In the pseudo-Dirac case, the Majorana terms are small and maximal mixing of active and sterile states occurs, which is generally excluded for solar neutrinos. In a ``Diracian''  limit, the physical masses become pairwise degenerate and the neutrinos attain a Dirac signature. 
Members of a pair do not oscillate mutually so that their mixing can be undone, and the standard neutrino model follows as a limit. While two Majorana phases become physical Dirac phases and three extra mass parameters occur,  a better description of data is offered. Oscillation problems are worked out in vacuum and in matter. With  lepton number --1 assigned to the sterile neutrinos, the model still violates lepton number conservation and allows very feeble neutrinoless double beta decay. 
It  supports a sterile neutrino interpretation of  Earth-traversing ultra high energy events detected by ANITA.
}



\begin{document}

\section{Introduction}
\label{sec:intro}

Thus far the Large Hadron Collider (LHC) has not produced evidence for physics beyond the standard model (BSM).
But the neutrino sector must involve BSM because neutrinos have mass. Indeed, the 2015 Noble prize in physics was awarded to 
T. Kajita  and A. B. McDonald  ``for the discovery of neutrino oscillations which show that neutrinos have mass'' \cite{Nobel2015}. 

The standard neutrino model (\SnuM) with its three Majorana neutrinos 
has measured values for the mass-squared differences, the mixing angles $\theta_{12}$, $\theta_{23}$ and $\theta_{13}$ and the weak Dirac phase $\delta$.
But the absolute mass scale, the order of the hierarchy, normal or inverted, and the Majorana phases are unknown.
There is stress in the fit to the standard solar model\cite{VinyolesNewGeneration2017}; 
 there is a  reactor neutrino anomaly \cite{PhysRevD.83.073006,dentler2017sterile}; 
MiniBooNE finds 4.5$\sigma$ evidence for a sterile neutrino\cite{aguilar2018significant}, 
while MINOS/MINOS+  does not \cite{adamson2019search}.
At present, there is no definite conclusion about the existence of an eV sterile neutrino \cite{boser2019status}.

There is also input from cosmology. 
From the lensing of background galaxies by the large, reasonably relaxed galaxy clusters 
Abell 1689\cite{nieuwenhuizen2009non,nieuwenhuizen2013observations,nieuwenhuizen2016dirac} and Abell 
1835\cite{nieuwenhuizen2017subjecting} there is indication for 3 active and 3 sterile neutrinos with common mass of 
 1.5--1.9 eV, which act as the cluster dark matter.
 We  shall not dwell here into the many questions raised by and counter-evidence to that possibility, but refer to the discussion and cited articles in these references.
Be it as it may,  the 3+3 case puts forward to consider the minimal extension of the standard model (SM) in the neutrino sector. By default, this accepts all SM physics 
without extension in the Higgs, gauge, quark and charged lepton sectors.  Gauge invariance then forbids the presence of a `left handed' Majorana 
mass matrix between the left handed active neutrinos, so that there {\it must} be a Dirac mass matrix to give them mass.
As such a term mixes left and right handed fields, this presupposes the existence of 3 {\it right handed} neutrinos,
also called {\it sterile}, i. e., not involved in elementary particle processes\cite{giunti2007fundamentals}. 
For that reason, they are allowed to have a  mutual `right handed' Majorana mass matrix.
In order to make up for half of the cluster dark matter,  sterile neutrinos have to be generated  in the early cosmos by oscillation of active ones.
This is only possible when the Dirac mass matrix is accompanied by a non-trivial right handed Majorana mass matrix.

In the pseudo-Dirac limit, the right handed Majorana masses are much smaller than the eigenvalues of the Dirac mass matrix.
The maximal mixing of the resulting pseudo-Dirac neutrinos implies that in principle half of the emitted solar neutrinos has become sterile here on Earth, 
and thus unobservable\voetnoot{See section \ref{intermezzo} for details.}; this is ruled out by the standard solar model\cite{VinyolesNewGeneration2017}.
Hence the pseudo-Dirac case is often considered to be ruled out. We intend to show, however, that there is a way out of this conundrum,
so as to faithfully include neutrino mass in the SM without changing its high energy sector.

While excellent studies such as \cite{gonzalez2008phenomenology,giunti2007fundamentals,lesgourgues2013neutrino}
discuss the theory for general number $N_s$ of sterile neutrinos, 
we shall  work out the case $N_s=3$ in a nontrivial limit where the 6 Majorana neutrinos combine into 3 Dirac 
neutrinos so that the maximal mixing is harmless and can be circumvented. We call them Diracian neutrinos, i. e., Dirac neutrinos in a model with 
both Dirac and Majorana masses.
In section 2 we treat the theory and in section 3 we consider various applications. We close with a summary.

\section{The Lagrangian for active plus sterile neutrinos}

In this section we concentrate on the neutrino sector of the SM.
For completeness we present the full Lagrangian in Appendix B. 

\subsection{Active neutrinos only}

We start from the SM Lagrangian where the $e$, $\mu$ and $\tau$ fields are diagonal in the mass basis.
Left handed neutrinos and right handed antineutrinos exist are called ``active neutrinos'' since they participate in  the weak 
interactions\voetnoot{Left and right handedness refers to the chirality; see Appendix A .}.
Additional neutrinos are not involved in them and called sterile.
If only active ones exist, they are Majorana particles. Their mass term involves the quantized left handed 
fermionic flavor fields $\nu_{e L},\nu_{\mu L},\nu_{\tau L}$, 
\BEQ\label{LmML}
\cL^{M}_{mL}=\half \sum_{\alpha,\beta=e,\mu,\tau} \nu_{\alpha L}^T\cCdag (M^{M}_L)_{\alpha\beta}\nu_{\beta L}+\hc
\EEQ
where  $\cC$ is the charge conjugation matrix, $T$ denotes transposition,  $\dagger$ Hermitian conjugation, and $\hc$ Hermitian conjugated terms.
$ M^M_L$ is called the left handed Majorana mass matrix. In the SM gauge invariance forces $ M^M_L$ to vanish\cite{giunti2007fundamentals}; 
if it is present, it must originate from high energy BSM, such as Weinberg's dimension-5 operator. 
Considering new physics only in the neutrino sector, we neglect $ M^M_L$.

\subsection{The Dirac and Majorana mass matrices}

In absence of $M^M_L$, the only possibility to give mass to the active neutrinos is by a Dirac mass matrix.
Since that involves products of left and right handed fields, this presupposes the existence of $N_s\ge 3$ sterile neutrinos,
that must be right handed and represented by quantized fermionic fields $\nu_{iR}$, 
\BEQ\label{LDirac}
\cL_m^{D}=-\sum_{\alpha=e,\mu,\tau} \sum_{i=1}^{N_s}
\Big (
\overline{\nu_{\alpha L}}M^D_{\alpha i}\nu_{iR}+\overline{\nu_{iR}}M^{D\,\dagger}_{i\alpha }\nu_{\alpha L}\Big),
\EEQ
where the Dirac mass matrix $M^D$ is a complex $3\times N_s$ matrix.
The sterile fields do not enter the weak interactions; they are singlets under the 
U(1)$_Y\times$SU(2)$_L\times$ SU(3)$_C$ gauge groups of the SM and affect neither gauge invariance, anomalies nor renormalization.
Hence they preserve its full functioning while accounting for neutrino masses.
Moreover, the sterile fields may have a mutual mass term like Eq. (\ref{LmML}), 
\BEQ\label{LmMR}
\cL^{M}_{mR}\!=\!\half \sum_{i,j=1}^{N_s}\Big(\nu_{iR}^T\cCdag M^{M\dagger}_{R,ij} \, \nu_{jR}
+ (\nu_{iR}^\cc)^ T\,\cCdag M^M_{R,ij}\,\nu_{j R}^\cc\Big),
\EEQ
where the right handed Majorana mass matrix $M^M_R$ is symmetric and complex valued, and where  
$\nu_{i R}^\cc$ is the charge conjugate of $\nu_{i R}$,
\BEQ
\nu_{i R}^\cc=\cC\, \overline{\nu_{i R}}\,^T=\cC\,(\gamma^{\,0})^T (\nu_{iR}^\dagger)^T=-\gamma^{\,0}\cC (\nu_{iR}^\dagger)^T .
\EEQ
While $\nu_{iR}$ is a right handed field, $\nu_{iR}^\cc$ is left handed (see Appendix A for properties of $\gamma$ and $\cC$ matrices).

The kinetic term has a common form for all species\cite{giunti2007fundamentals},
\BEQ
\cL_k=\sum_{\alpha=e,\mu,\tau}\overline{\nu_{\alpha L}}\ci\overleftrightarrow{\slashed{\p}}\nu_{\alpha L}+
\sum_{i=1}^{N_s}\overline{\nu_{iR}}\ci\overleftrightarrow{\slashed{\p}}\nu_{ iR},
\EEQ
where the slash denotes contraction with $\gamma$ matrices, and the partial derivatives acting as
\BEQ
\overleftrightarrow{\slashed{\p}}=\sum_{\mu=0}^3\gamma^{\,\mu} \overleftrightarrow{\p_\mu},\qquad 
\overleftrightarrow{\p_\mu}=\frac{\overrightarrow{\p_\mu}-\overleftarrow{\p_\mu}   }  {2},
\qquad 
\overline{a}\,\slashed{\p}b\equiv 
\frac{1}{2}\sum_{\mu=0}^3\Big(\overline{a}\gamma^\mu\frac{\p b}{\p x^{\,\mu} }- \frac{\p{\overline{a}}}{\p x^{\,\mu}}\,\gamma^{\mu}b\,\Big).
\EEQ

\subsection{The general mass matrix for 3 sterile neutrinos}

Though the number of right handed neutrinos is not fixed in principle, the case $N_s=3$ has, if not a practical 
value\cite{nieuwenhuizen2009non,nieuwenhuizen2013observations,nieuwenhuizen2016dirac,nieuwenhuizen2017subjecting}, 
at least an esthetic one: for each left handed neutrino there is a right handed one, in the way it occurs for charged leptons and quarks.
The three families of \underline{a}ctive left and \underline{s}terile right handed neutrinos have the \underline{f}lavor 3 vectors\voetnoot{In our case 
$N_s=3$ one may be tempted to denote $(\nu_{1R},\nu_{2R},\nu_{3R})$ as $(\nu_{eR},\nu_{\mu R},\nu_{\tau R})$.}
\BEQ\label{nufL}
\bm{\nu}_{fL}\equiv
 \bm{\nu}_{aL}=(\nu_{e L},\nu_{\mu L},\nu_{\tau L})^T , 
 \qquad \bm{\nu}_{fR}\equiv \bm{\nu}_{sR}=(\nu_{1R},\nu_{2R},\nu_{3R})^T .
\EEQ
With the combined left handed flavor vector 
\BEQ \label{NfL}
\bN_{fL}=(\bm{\nu}_{fL}^T,\bm{\nu}_{fR}^{\cc\,T})^T 
=( \bm{\nu}_{aL}^T, \bm{\nu}_{sR}^{c\,T})^T  ,
\EEQ
the above mass Lagrangians combine into
\BEQ\label{LMass}
\cL_m=\half \bN_{fL}^T\cCdag M^\DpM \bN_{fL}+\hc 
\EEQ
In general, the mass matrix consists of four $3\times 3$ blocks,
\BEQ \label{blockM}
M^\DpM=\begin{pmatrix}M_L^M & M^D{}^T \\ M^D & M_R^M \end{pmatrix} .
\EEQ
As stated, we take $M_L^M=0$. 
{ In the (``standard'', ``pure'' or ``trivial'') Dirac limit also $M_R^M=0$. 
For pseudo-Dirac neutrinos $M^M_R$ will be small with respect to $M^D$, or, more precisely, small with respect to the variation in the eigenvalues of $M^D$.
Though we consider general $M^D$, we are inspired by the case of galaxy cluster lensing where it has nearly equal eigenvalues with 
central value 1.5--1.9 eV \cite{nieuwenhuizen2009non,nieuwenhuizen2013observations,nieuwenhuizen2016dirac,nieuwenhuizen2017subjecting};
 in section \ref{massestimates} we shall show that the entries of $M^M_R$ then typically lie well below 1 meV.}

\subsection{Intermezzo: One neutrino family}
\label{intermezzo}

In case of one family the flavor vector is $\bN_{fL} =(\nu_{L},\nu_{R}^c)^T$. The entries of Eq. (\ref{blockM}) are scalars, so 
\BEQ \label{fam1}
M^\DpM=\begin{pmatrix}0 & \mb \\ \mb & \mub \end{pmatrix} .
\EEQ
Its eigenvalues are
\BEQ
\lambda_{1,2}=\frac{1}{2}\mub\mp \sqrt{\mb^2+\frac{1}{4}\mub^2}.
\EEQ
The physical masses are their absolute values\cite{giunti2007fundamentals}. For $\mub$ nonnegative, this leads to
\BEQ\hspace{-5mm}
m_{1}=\sqrt{\mb^2+\frac{1}{4}\mub^2}-\frac{1}{2}\mub, \quad m_{2}=\sqrt{\mb^2+\frac{1}{4}\mub^2}+\frac{1}{2}\mub.
\EEQ
The corresponding eigenvectors are
\BEQ &&
\ve^{(1)}
=\frac{1}{\sqrt{\mb^2+m_1^2}}\begin{pmatrix}\mb \\ -m_1\end{pmatrix} 
=\frac{1}{\sqrt{\mb^2+m_2^2}}\begin{pmatrix}m_2\\ -\mb\end{pmatrix} 
,\qquad 
\ve^{(2)}=\frac{1}{\sqrt{\mb^2+m_1^2}}\begin{pmatrix}m_1\\ \mb\end{pmatrix} .
\qquad 
\EEQ
For small $\mub$ these are 45$^\circ$ rotations, i. e., maximal mixing of the active and sterile basis vectors.
Formally we may undo the rotations over 45$^\circ$, by considering
\BEQ\label{eaes1fam} &&
\ve^\ta=\frac{\ve^{(1)}+\ve^{(2)}}{\sqt}\approx \begin{pmatrix}1 \\ {\mub}/{4\mb}\end{pmatrix},\qquad 
\ve^\ts=\frac{\ve^{(2)}-\ve^{(1)}}{\sqt}\approx \begin{pmatrix}- {\mub}/{4\mb} \\ 1\end{pmatrix} ,
\EEQ
where the approximations are to first order in $\mub$, the pseudo Dirac regime. The first vector, $\ve^\ta$, has its main weight on the first component, 
so it is mainly active, which we indicate by the tilde on $a$. 
The second one, $\ve^\ts$, is mainly sterile. But unless $\mub=0$, the masses $m_{1,2}$ are different, so that $\ve^\ta$ and $\ve^\ts$ 
are not eigenvectors and have no physical meaning. In fact, the mass squares have the difference
\BEQ 
\Delta m^2_{21}=m_2^2-m_1^2=2 \mub\sqrt{\mb^2+\frac{1}{4}\mub^2}\approx 2\mub \mb.
\EEQ
An initially active state,
\BEQ
|\nu_a(0)\rangle=\begin{pmatrix}1\\ 0 \end{pmatrix} =\frac{\mb \ve^{(1)}+m_1\ve^{(2)}}{\sqrt{\mb^2+m_1^2}},
\EEQ
with momentum $p$ will at time $t$ have oscillated into
\BEQ
|\nu_a(t)\rangle=\frac{\mb \ve^{(1)}e^{-\ci E_1t}+m_1 \ve^{(2)}e^{-\ci E_2t}}{\sqrt{\mb^2+m_1^2}},
\EEQ
where $E_{1,2}=\sqrt{p^2+m_{1,2}^2}$. The occurrence probability is
\BEQ\label{Paa1fam}
P_{aa}(t)&=&|\langle\nu_a(0)|\nu_a(t)\rangle|^2=\frac{\mb^4+m_1^4+2\mb^2m_1^2\cos{\Delta E  t}}{(\mb^2+m_1^2)^2} 
\approx \frac{1+\cos{\Delta E  t}}{2},
\EEQ
where for $p\gg \mb$
\BEQ
\Delta E\equiv E_2-E_1\approx \frac{\Delta m_{21}^2}{2p}\approx \frac{\mub\mb}{p}.
\EEQ
In practice there will not be a pure initial state but some wave packet\cite{giunti2007fundamentals}. 
For $t\gg \hbar/\Delta E$ the cosine in Eq. (\ref{Paa1fam}) will average out, so that the fraction of observable neutrinos is approximately $\half$. 
In plain terms: for $t$ large enough, half of the neutrinos are sterile and thus unobservable.
For the solar neutrino problem the one-family approximation happens to work quite well\cite{smirnov2016solar}
and the detection rates are well established.
Hence for the pseudo Dirac model it would mean that twice as many neutrinos should be emitted as in the standard solar model. 
The corresponding doubling of heat generated by nuclear reactions is ruled out by the measurements of the solar luminosity,
so the case is rarely discussed.

Only in the pure Dirac case, i. e., with  Majorana mass $\mub=0$, the oscillations will not take place, since $m_{1,2}=\mb$ and $\Delta E=0$. 
When starting from an initial active state $\nu_a(0)$, it now  equals $\ve^\ta$, and this can be taken as eigenstate. 
The sterile state will merely be a spectator, ``just sitting there and wasting its time''. This can be generalized to three families.
If one would follow the Franciscan William of Ockham (Occam's razor), it would be preferable for active neutrinos to be Majorana rather than 
Dirac with unobservable right handed partners. 

The  \SnuM \,  differs from the neutrino sector in the SM by accounting for finite masses of its 3 Majorana neutrinos.
Below we discuss a ``Diracian'' setup in which the sterile fields become physical, namely partly active, and the active fields partly sterile,
even though the mass eigenstates have Dirac signature in vacuum.

\subsection{Diagonalization of the Dirac mass matrix}

We return to the 3 family case and its total mass matrix Eq. (\ref{blockM}) with $M_L^M=0$.
We notice that any $3\times3$ unitary matrix $U$ can be decomposed as a product of 5 standard ones,
\BEQ\label{Udecomp}
U=D'U^\DpM,\quad U^\DpM=U^DD^M,\qquad 
 U^D=U_1U_2U_3,\qquad D^M=\diag(e^{\ci \eta_1},e^{\ci \eta_2},e^{\ci \eta_3}).
\EEQ
The diagonal matrix $D^M$ is called the Majorana phase matrix. 
Likewise we denote the diagonal phase matrix\voetnoot{For $U$ in Eq.  (\ref{Udecomp}) only 5 of the $\eta_i$ and $\eta'_i$ are needed; 
this can be seen by factoring out $e^{\ci \eta_1}$ from $D^M$ and $e^{\ci \eta_1'}$ from $D'$ and setting $\eta_1\to\eta_1-\eta_1'$. 
Both sides of Eq. (\ref{Udecomp}) thus involve 9 free parameters.}  $D'$ by $\diag(e^{\ci \eta'_1},e^{\ci \eta'_2},e^{\ci \eta'_3}$). 
The matrix $U^D$ is the product of 
\BEQ\label{U1U2U3}
&&
U_1=\begin{pmatrix}  1 & 0 & 0\\ 0& c_{1}& s_{1} \\ 0& -s_{1}& c_{1} \end{pmatrix} , \quad
  U_2=\begin{pmatrix}  c_{2} & 0 & s_{2}e^{-\ci\delta} \\ 0& 1 & 0 \\  -s_{2}e^{\ci \delta} & 0 & c_{2} \end{pmatrix},\quad 
  U_3=\begin{pmatrix}  c_{3} & s_{3} & 0\\ -s_{3}& c_{3}& 0\\ 0& 0&1 \end{pmatrix} , 
\EEQ
where $c_i=\cos\theta_i$, $s_i=\sin\theta_i$ where the angles $\theta_i$ are termed in standard notation $\theta_1=\theta_{23}$, 
$\theta_2=\theta_{13}$ and $\theta_3=\theta_{12}$. The Dirac phase $\delta$ is also called weak CP violation phase.

The complex valued Dirac mass matrix $M^D$ can be diagonalized by two unitary matrices 
 of the form (\ref{Udecomp}), viz. $U_{L}=D_L'U_L^DD_L^M$ and $U_R=D_R'U^D_RD_R^M$. The result reads
 \BEQ
\label{MDdiag}
M^D=U_R^{T\dagger}M^d U_L^\dagger,\qquad  
M^d=\diag(\mbe ,\mbt ,\mbd ), 
\EEQ
with the real positive $\mbi$\voetnoot{We denote Dirac mass eigenvalues by $\mb_i$ to distinguish them from the physical masses $m_i$, 
the eigenvalues in absolute value of the total mass matrix.}{$^,$}\voetnoot{While the left hand side of Eq. (\ref{MDdiag}) has 9 complex or 18 real parameters,  
the right hand side has $9 + 3+9$; but since $M^d$ is diagonal, the diagonal matrices $D_L^{M\dagger}=D_L^{M\ast}$ and $D_R^{M\ast}$ only act as a product. 
Hence it is allowed to fix $D^M_R$ before solving $D^M_L$, see below Eq. (\ref{myMm=}). The number of parameters available for the diagonalization is then still 18.\label{fn5}}.
We identify $U_L^D$ with the  {PMNS} mixing matrix $U^D=U_1U_2U_3$ and $D_L^M$ with the Majorana matrix 
$D^M$ employed in literature.

To connect the transformation (\ref{MDdiag}) to $M^{\DpM}$, we introduce the $6\times6$ unitary matrix
\BEQ \label{ULR=}
U_{LR}=\begin{pmatrix} U_L&0\\0& U_R\end{pmatrix} ,
\EEQ
and define, using that $M^{d\,T}=M^d$ since it is diagonal, 
\BEQ
\label{MDpM=}
\cM =U_{LR}^T\, M^{\DpM} U_{LR}=\begin{pmatrix} 0 & M^d \\  M^d & \mM  \end{pmatrix} ,
\EEQ
New active and sterile fields $\bn_{aL}=U_L^\dagger\bm{\nu}_{fL}$,  $\bn_{sR}=U_R^T\bm{\nu}_{fR}$, merged as 
\BEQ\label{nL=}
\bn_L=(n_{aL}^1,n_{aL}^2,n_{aL}^3,n_{sR}^{1c},n_{sR}^{2c},n_{sR}^{3c})^T,
\EEQ express (\ref{NfL}) as
\BEQ
\label{NfL2nL}
\quad \bN_{fL}=(\bm{\nu}_{fL}^T,\bm{\nu}_{fR}^{\cc\,T})^T =U_{LR} \bn_L,\quad \bn_L=U_{LR}^\dagger {\bf N}_{fL}.
\EEQ
With these steps the right handed Majorana mass matrix transforms into
\BEQ\label{myMm=}
 \mM = U_R^T\MM  U_R . 
\EEQ
Like $\MM $, it is complex symmetric, but since $U_R$ was needed to diagonalize $M^D$, it will in general not result in a diagonal $\mM$.
With the decomposition $U_R=D_{R}U_R^DD_{R}^M$ as in Eq. (\ref{Udecomp}), one can, however, use the phases in $D_R^M$
to make the off-diagonal elements of $\mM $ real and nonnegative$^\text{\ref{fn5},}$\voetnoot{Actually, for $n$ lepton families there are $\half n(n-1)$ independent
complex valued off-diagonal elements and $n$ Majorana phases, so making all off-diagonal elements real and nonnegative is possible for $n=3$ or 2. }.

We denote the diagonal elements of the Majorana matrix $\mM $ by $\mubi$, that may still be complex, and the real positive off-diagonal elements by $\mu_i$.
The right handed Majorana mass matrix $M^N$ then takes the form
\BEQ \label{cMLR=}
M^N=\begin{pmatrix} 
 \mube & \mu_{3} & \mu_2  \\
\mu_3& \mubt & \mu_1  \\
\mu_2&  \mu_1& \mubd \\
\end{pmatrix} , \hspace{2mm}
\EEQ  
so that the total mass matrix $\cM$ reads
\BEQ
\label{cMmatrix}
\cM=\begin{pmatrix} 
0 & 0 & 0 & \mDe  &0 &0 \\
0&0 & 0 & 0 & \mDt  &0 \\
0&0 & 0 & 0 &0& \mDd  \\
\mDe &0&0 & \mube & \mu_{3} & \mu_2  \\
0&\mDt &0&\mu_3& \mubt & \mu_1  \\
0&0& \mDd &\mu_2&  \mu_1& \mubd \\
\end{pmatrix} .
\EEQ
Except in the pure Dirac limit where $\mu_i=\mubi=0$, 
the $\bn_L$ are not rotations of mass eigenstates.

\vspace{1mm}

\subsection{Diracian limit} 

For reasons explained above, we wish to achieve pairwise degeneracies in the masses.
The standard Dirac limit,  just taking $\mu_i$ and $\mubi\to 0$, is a trivial way to achieve this; 
we shall, however, need finite values for them and design the more subtle ``Diracian'' limit.

To start, we notice that the eigenvalues of the mass matrix (\ref{cMLR=}) follow from ${\rm det}(\cM-\lambda I)=0$, where 
\BEQ \label{veryfulldet}
&& {\rm det}(\cM -\lambda I)=
\\ &&(\lambda^2\mmin \mbe ^2)(\lambda^2\mmin \mbt ^2) (\lambda^2\mmin \mbd ^2)   
- \,(\mube \pplus  \mubt \pplus  \mubd) \lambda^5
-(\mu_1^2 \pplus \mu_2^2 \pplus \mu_3^2 \mmin \mube\mubt\mmin \mubt\mubd\mmin \mubd\mube)\lambda^4 \,
\nn \\&& 
+  [\mDe ^2(\mubt\pplus \mubd) \pplus \mDt ^2(\mubd\pplus  \mube)\pplus \mDd ^2 (\mube\pplus \mubt) 
\pplus\mu_1^2 \mube\pplus  \mu_2^2 \mubt\pplus \mu_3^2 \mubd\mmin 2\mu_1\mu_2\mu_3\mmin \mube\mubt\mubd]\lambda^3
 \nn\\&& 
 \pplus [\mDe ^2( \mu_1^2 \mmin  \mubt\mubd)\pplus \mDt ^2 (\mu_2^2\mmin \mubd\mube) \pplus \mDd ^2 (\mu_3^2 \mmin  \mube\mubt)]\lambda^2  
 \mmin (\mDe ^2 \mDt ^2 \mubd\pplus \mDt ^2 \mDd ^2 \mube\pplus \mDd ^2 \mDe ^2 \mubt)\lambda .  \nn
\EEQ
The criterion to get pairwise degeneracies in the eigenvalues (up to signs), is simply that the odd powers in $\lambda$ vanish.
Let us denote
\BEQ
&& 
 \Deltae=\mb _2^2-\mb _3^2,\quad
\Deltat=\mb _3^2-\mb _1^2,\quad
\Deltad=\mb _1^2-\mb _2^2,\quad 
\nn\\ && 
\Mbe=\frac{\mbt\mbd}{\mbe},\qquad 
\Mbt=\frac{\mbd\mbe}{\mbt},\qquad 
\Mbd=\frac{\mbe\mbt}{\mbd}, 
\EEQ
and express the $\mubi$ in a common dimensionless parameter $\ub$ through
\BEQ \label{mubi=}
 \mubi =\frac{ \Deltai }{ \Mbi} \ub .
\EEQ 
The relations $\sum _i\mbi^2\Deltai=\sum _i\Deltai=0$ make the coefficients of $\lambda^5$ and $\lambda^1$ of Eq. (\ref{veryfulldet}) vanish, respectively.
To condense further notation, we express the $\mu_i$ into dimensionless non-negative parameters $u_i$, 
\BEQ \label{muiui}
\mu_i=\sqrt{|\Deltae \Deltat \Deltad|}\frac{u_i}{\mb_i\sqrt{|\Deltai|}}.
\EEQ
For normal ordering of the $\mbi$ (notice that these are Dirac masses, not the physical masses),
 $\mbe <\mbt <\mbd $ implies $\Deltae<0$, $\Deltat>0$, $\Deltad<0$, hence $\Deltae\Deltat\Deltad>0$;
this is also the case for the inverted ordering  $\mbd <\mbe <\mbt $ whence $\Deltae>0$, $\Deltat<0$, $\Deltad<0$. It thus holds that  
 \BEQ
\mu_1\mu_2\mu_3=\frac{\Deltae\Deltat\Deltad}{\mbe \mbt \mbd }u_1u_2u_3 .
\EEQ
Equating the $\lambda^3$ coefficient of Eq. (\ref{veryfulldet}) to zero requires
\BEQ \hspace{-3mm}
\ub^3 -  (1 + u^2) \ub+2 u_1 u_2 u_3=0,\quad u^2\equiv \frac{\Deltae}{|\Deltae|} u_1^2 +  \frac{\Deltat}{|\Deltat|} u_2^2 + \frac{\Deltad}{|\Deltad|}  u_3^2=
\frac{\Deltae}{|\Deltae|} \big(u_1^2-u_2^2\big)  -  u_3^2.
  \EEQ
This cubic equation has the solutions for $n=-1,0,1$ and positive or negative $u^2$
\BEQ &&
\hspace{-3mm} 
\label{mub=}
\ub=\frac{-\ci\,} { \sqrt{3} }\Big[e^{2\pi \! \ci n/3} \,u_+^{1/3}-e^{-2\pi \! \ci n/3} (1+u^2)\,u_+^{-1/3}\Big]   ,  \\ &&  \hspace{-3mm} 
u_+=\sqrt{D} \pplus\ci \sqrt{27}u_1 u_2 u_3, \hspace{2mm} D=  (1 \pplus u^2)^3 \mmin 27u_1^2u_2^2u_3^2. \hspace{2mm} \nn
\EEQ
We restrict ourselves to real solutions; there is always one. Then the matrix $\cM$ is real valued.
All solutions are real when $D> 0$, which occurs in particular when the $u_{i}$ are small,
i.e.,  in the pseudo-Dirac case.
Then there exist the large solutions $n=\pm1$ with $\ub\approx \pm 1$, which in both cases leads to the eigenvalues $\lambda_i^\pm\approx\pm M_i$ for $i=1,2,3$.
For small $u_i$ the $n=0$ solution has a small $\ub$ and $\mubi$, viz.
\BEQ \label{mubmui=}
&&
\ub\approx2u_1u_2u_3=2\frac{ \mu_1\mu_2\mu_3 }{\Deltae\Deltat\Deltad}\mbe \mbt \mbd,\quad
\mubi\approx 2\frac{\mu_1\mu_2\mu_3}{\Deltae\Deltat\Deltad}\mb_i^2\Deltai =2u_1u_2u_3\frac{\Deltai}{M_i} .
\EEQ

The Diracian limit, defined by Eqs. (\ref{mubi=}), (\ref{muiui}) and (\ref{mub=}), reduces Eq. (\refl{veryfulldet}) to a cubic polynomial in $\lambda^2$,
\BEQ 
\label{fulldet} 
&& (\lambda^2-\mbe ^2)(\lambda^2-\mbt ^2) (\lambda^2-\mbd ^2) \\&&
+\lambda^2\Deltae \Deltat \Deltad
\Big[   \frac{(\lambda^2-\mbe^2)(\ub^2 - u_1^2)}{\mbe^2\Deltae}+ \frac{(\lambda^2-\mbt ^2)(\ub^2 -  u_2^2)}{\mbt ^2\Deltat}
+ \frac{(\lambda^2-\mbd^2)(\ub^2 - u_3^2)}{\mbd^2\Deltad} \Big] =0 . \nn
\EEQ
 Its analytical roots are intricate, but they are easily calculated numerically.
Denoting them as $m_i^2$, the squares of the physical masses, the eigenvalues of $\cM$ are
$\lambda_{2i-1}=-\, m_i$ and $\lambda_{2i}=+ \,m_i>0$ for $i=1,2,3$. 
From  det $\cM=-\mbe^2\mbt^2\mbd^2$ it holds that $m_1m_2m_3=\mb_1\mb_2\mb_3$.
The eigenvectors are set by
\BEQ\label{cMdecomp}
\sum_{k=1}^6\cM_{jk} e_k^{(i)}=\lambda_i e_j^{(i)},\qquad (i=1,\cdots,6) .
\EEQ
and they are real and orthonormal.  They can be expressed as
\BEQ\label{e2im12i}
\ve^{(2i-1)}=\frac{\ve^{(i\ta)}-\ve^{(i\ts)}}{\sqrt{2}} , \quad  \ve^{(2i)}=\frac{\ve^{(i\ta)}+\ve^{(i\ts)}}{\sqrt{2}} .
\EEQ
with orthonormal $ \ve^{(i\ta)}$ and $\ve^{(i\ts)}$ for $i=1,2,3$. For small  $\mu_i$ and $\mubi$ 
the $\ve^{(i\ta)}$ and $\ve^{(i\ts)}$ read to first order\voetnoot{Notice that the $i$ and $i+3$ components of $\ve^{(i\ta)}$ and $\ve^{(i\ts)}$ 
stem with the one-family case Eq. (\ref{eaes1fam}).}
\BEQ \label{eigvcs} \label{eigvA} 
&& \hspace{-0mm}
\ve^{(1\ta)}=\Big(1,0,0,\frac{\mube}{4\mbe } ,\mDe\frac{ \mu_3}{\Deltad} ,-\mDe\frac{ \mu_2}{\Deltat}\Big)^T , 
\quad \ve^{(1\ts)}= \Big(-\frac{\mube}{4\mbe },  \mDt\frac{\mu_3}{\Deltad}  ,-\mDd\frac{ \mu_2}{\Deltat}, 1,0,0\Big)^T , 
\nn\\ && \hspace{-0mm} \label{eigvS}
\ve^{(2\ta)}=\Big( 0,1,0,-\mDt\frac{ \mu_3}{\Deltad},\frac{\mubt}{4\mbt },\mDt\frac{ \mu_1}{\Deltae } \Big)^T , 
\quad
\ve^{(2\ts)}= \Big(-\mDe \frac{\mu_3}{\Deltad} ,-\frac{\mubt}{4\mbt },\mDd\frac{ \mu_1}{\Deltae },0,1,0 \Big)^T , \,\, 
\\ && \hspace{-0mm}
\ve^{(3\ta)}= \Big(0,0,1, \mDd\frac{ \mu_2}{\Deltat} , - \mDd\frac{ \mu_1}{\Deltae } ,\frac{\mubd}{4\mbd }\Big)^T ,  
\quad
\ve^{(3\ts)}= \Big(\,\, \mDe\frac{ \mu_2}{\Deltat} , - \mDt\frac{ \mu_1}{\Deltae }  ,-\frac{\mubd}{4\mbd },0,0,1\Big)^T .   \nn 
\EEQ
With the first three components of these vectors relating to active neutrinos and the last three to sterile ones, it is seen that for
small $\mu_i$ and $\mubi$ the $\ve^{(i\ta)}$ are {\it mainly active} and the $\ve^{(i\ts)}$ {\it mainly sterile}, which we indicate by the tildes.
The $\ve^{(2i-1)}$ and $\ve^{(2i)}$ are $45^\circ$ rotations of the $\ve^{(i\ta)}$ and $\ve^{(i\ts)}$, which is maximal mixing.
In the standard Dirac limit, it is customary to work with Dirac states and not with Majorana states. Likewise,
in our Diracian limit the mass degeneracies allow the rotations to be circumvented by working with the $\ve^{(i\ta)}$ and $\ve^{(i\ts)}$ themselves.
Indeed, there holds the exact decomposition 
\BEQ&&
\cM=
\sum_{i=1}^6\lambda_i\ve^{(i)}\ve^{(i)\,T}=
\sum_{i=1}^3 m_i\Big[\ve^{(i\ta)}\ve^{(i\ts)\,T}+\ve^{(i\ts)}\ve^{(i\ta)\,T}\Big] .
\EEQ
These steps allow to retrieve the standard Dirac expressions in the limit where the Majorana masses $\mu_i$, $\bar\mu_i$ vanish,
whence the $\ve^{(i\ta)}$ and $\ve^{(i\ts)}$ become purely active and purely sterile states, respectively.

For neutrino oscillation probabilities in vacuum (see subsections \ref{intermezzo} and  \ref{nu-osc}) one needs the eigenvalues $m_i^2$ 
and eigenvectors $\ve^{(i\ta)}$ and $\ve^{(i\ts)}$ of $\cM^2$, 
\BEQ
\cM^2=\sum_{i=1}^3 m_i^2\Big[\ve^{(i\ta)}\ve^{(i\ta)\,T}+\ve^{(i\ts)}\ve^{(i\ts)\,T}\Big].
\EEQ

In terms of $\bn_L$ defined above Eq. (\ref{NfL2nL}) and related there to the flavor states (\ref{NfL}),
the fields for the mass eigenstates are
\BEQ\label{numanums}  \hspace{-5mm}
\nu_{\ta L} ^{i}=\sum_{j=1}^6e^{(i\ta)}_j n_{jL}, \qquad \nu_{\ts R}^{i}=\sum_{j=1}^6e^{(i\ts)}_j n_{jL}^{\cc},\quad (i=1,2,3) .
\EEQ
Here $j=1,2,3$ label active fields and $j=4,5,6$ sterile ones. 
Hence the fields $\nu_{\ta L} ^{i}$ annihilate chiral left handed, {\it mainly} active neutrinos and create similar right handed antineutrinos, 
while the $\nu_{\ts R}^{i}$ annihilate chiral right handed, {\it mainly} sterile neutrinos and create similar left handed antineutrinos.

 The mass term of the 6 Majorana fields now takes the form of 3 Dirac terms, 
\BEQ
\hspace{-1mm}
 \cL_m
 \hspace{-0mm}
 &=&\!\half\sum_{i=1}^3 m_i\Big[\nu_{\ta L} ^{i\,T}  \cCdag \nu_{\ts R}^{i\cc}+(\nu_{\ts R}^{i\cc})^T  \cCdag \nu_{\ta L} ^{i} \Big] +\hc 
=\half\sum_{i=1}^3 m_i\Big ({\nu_{\ta L} ^{i}}^{T}\overline{\nu_{\ts R}^i}^{\,T}-\overline{\nu_{\ts R}^i}\nu_{\ta L} ^{i}\Big)+\hc  \nn\\
&=& -\sum_{i=1}^3m_i\Big(\overline{\nu_{\ts R}^i}\nu_{\ta L}^i +\overline{\nu_{\ta L}^i}\nu_{\ts R}^i \Big)
=-\sum_{i=1}^3m_i\overline{\nu^i}\nu^i , \label{Hnum}
\EEQ
because fermion fields anticommute and left and right handed fields are orthogonal. The here introduced Dirac fields,
\BEQ
\nu^i =\nu_{\ta L} ^i+\nu_{\ts R}^i ,\qquad (i=1,2,3),
\EEQ
combine left and right handed chiral fields,  as usual.  They are the mass eigenstates. 
In this basis the Dirac-Majorana neutrino Lagrangian is a sum of Dirac terms,
\BEQ \label{cLDM=}
\cL=\sum_{i=1}^3\Big(\overline{\nu^i}\ci\overleftrightarrow{\slashed{\p}}\nu^{i}-m_i\overline{\nu^i}\nu^{i}\Big).
\EEQ

\subsection{Charged and neutral current}

Neutrinos also enter the currents coupled to the $W$ and $Z$ gauge bosons, which are part of the covariant derivatives in the Lagrangian,
see Eq. (\ref{DMnSM}) below. The $W$ boson couples to the \underline{c}harged weak \underline{c}urrent.  On the flavor basis it reads
\BEQ
\label{Jcc=} 
\cL_{CC}=-\frac{g}{2\sqt}\Big[W_\mu J^{\mu\,\dagger}_{CC}+W_\mu^\dagger J^{\,\mu}_{CC}\Big], \quad    
J^{\mu}_{CC}=2\sum_{\alpha=e,\mu,\tau}\overline{ \nu _{\alpha L}} \gamma^{\,\mu}\ell_{\alpha L} ,\quad
J^{\mu\,\dagger}_{CC}=2\sum_{\alpha=e,\mu,\tau}\overline{ \ell_{\alpha L}}\gamma^{\,\mu}\nu _{\alpha L} , 
\EEQ
with $g$ the weak coupling constant. The \underline{n}eutral weak \underline{c}urrent reads on the flavor basis
\BEQ  \label{Jnc=} \hspace{-5mm}
\cL_{NC}=-\frac{g}{2\cos\theta_w}Z_\mu J_{NC}^{\,\mu},\quad 
J_{NC}^{\,\mu}=\sum_{\alpha=e,\mu,\tau}\overline{\nu_{\alpha L}}\gamma^{\,\mu} \nu_{\alpha L} ,
\EEQ
with $\theta_w$ the weak or Weinberg angle. 

To express these in the mass eigenstates, we define
$\cA$ and $\cS$ as matrices consisting of the active components of the
6-component eigenvectors $\ve^{(i\ta)}$ and $\ve^{(i\ts)}$, respectively,
\BEQ\label{cAcSdef}
\cA_{ji}=e^{(i\ta)}_j,\quad \cS_{ji}=e^{(i\ts)}_j,\quad (j,i=1,2,3),
\EEQ
and, likewise,  $\cA^s$ and $\cS^s$ for the sterile components
\BEQ
\cA^s_{ji}=e^{(i\ta)}_{j+3},\quad \cS^s_{ji}=e^{(i\ts)}_{j+3},\quad (j,i=1,2,3).
\EEQ
From (\ref{eigvcs}) we read off that for small $\mu_i$ and $\mubi$
\BEQ \label{easv=}
&&
\cA_{ji}\approx\delta_{ij}, \qquad 
\cS_{ji}\approx -\delta_{ij}\frac{\mubi}{4\mbi }+ \sum_{k=1}^3\varepsilon_{ijk}\bar m_j \frac{\mu_k }{\Deltak} , \\&&
\cS^s_{ji}\approx\delta_{ij}, \qquad \,\,
\cA^s_{ji}\approx \,\,\, \delta_{ij}\frac{\mubi}{4\mbi }+ \sum_{k=1}^3\varepsilon_{ijk}\bar m_i \frac{\mu_k }{\Deltak} . 
\label{essv=}
\EEQ
From the orthonormality of the eigenvectors it follows that the real valued $3\times 3$ matrices $\cA$ and $\cS$  satisfy the unitarity relation
\BEQ  \label{AApSS=}
 \cA\cA^{T}+\cS\cS^{T}=\veen_{3\times3} , 
\qquad 
\EEQ
while $\cA^{T}\cA+\cS^{T}\cS\neq \veen_{3\times 3}$.

From Eqs.  (\ref{nufL}), (\ref{NfL}), (\ref{Udecomp}), (\ref{U1U2U3}), (\ref{ULR=}) and (\ref{NfL2nL}), and denoting $D^M\equiv D^M_L$, we have 
$\bm{\nu}_{fL}=D_L' U^D D^M\bn_{aL}$.  As shown below Eq. (\ref{Jnc=}), the diagonal phase matrix $D_L'$ can be absorbed in the fields.
Inverting Eq. (\ref{numanums}) leaves Eq. (\ref{cLDM=})  invariant and expresses the {\it flavor} 
eigenstates as superpositions of {\it mass} eigenstates $\bm{\nu}_{mL}=(\bm{\nu}_{\ta L},\bm{\nu_}{\ts R}^\cc)$.
In vector notation, and using $ \overline{\bm{\nu}_{\ts R}^c}=-\bm{\nu}_{\ts R}^T\cC^\dagger$, one has
\BEQ \label{nuf2num}
\bm{\nu}_{fL}
&=&U^\DpM(\cA \, \bm{\nu}_{\ta L}+\cS \, \bm{\nu}_{\ts R}^{\cc})  =A \, \bm{\nu}_{\ta L}+S \, \bm{\nu}_{\ts R}^{\cc} \, ,  \nn  \\
\overline{\bm{\nu}_{fL}} &=&(\overline{\bm{\nu}_{\ta L}}\cA^T+\overline{\bm{\nu}_{\ts R}^\cc}\cS^T)U^{\DpM\dagger} 
=\overline{\bm{\nu}_{\ta L}}A^\dagger+\overline{\bm{\nu}_{\ts R}^\cc} S^\dagger ,    \\ &=&
 (\overline{\bm{\nu}_{\ta L}}\cA^T-\bm{\nu}_{\ts R}^T\cC^\dagger\cS^T)U^{\DpM\dagger} 
=\overline{\bm{\nu}_{\ta L}}A^\dagger-\bm{\nu}_{\ts R}^T\cC^\dagger S^\dagger .
  \nn
\EEQ
Here $U^\DpM=U^DD^M$, with $U^D$ is the standard PMNS matrix,  see (\ref{Udecomp}), 
while $D^M=\diag(e^{\ci\eta_1},e^{\ci\eta_2},e^{\ci\eta_3})$ is  the Majorana matrix of the 3-neutrino problem; its $\eta_i$ 
are Dirac phases now\voetnoot{\label{fntMaj}A 
word on nomenclature: The Majorana phases in the matrix $D^M$ stem from the  3+0 \SnuM, without sterile neutrinos.
While they become physical Dirac phases  in the 3+3 \DMnSM, there appear no true 3+3 Majorana phases, so we propose to keep this name for them. 
 Hence the \DMnSM \, has 3 physical phases: 1 Dirac  and 2 ``Majorana'' phases. They all appear in the CP-invariance breaking part of the
neutrino oscillation probabilities, see Eq.  (\ref{DeltaPabCPv}).}. We also introduced
\BEQ
A=U^\DpM\cA,\quad 
S=U^\DpM\cS,\quad 
 AA^{\dagger}+SS^{\dagger}=\veen_{3\times3} .
\EEQ
The sterile field $\bm{\nu}_{sR}^\cc=(\nu_{sR}^{1\cc},\nu_{sR}^{2\cc},\nu_{sR}^{3\cc}$) similar  to (\ref{nuf2num}) reads 
\BEQ\label{nusnusa}
\bm{\nu}_{sR}^\cc=D_R'U^D_RD^M_R(\cA^s\bm{\nu}_\ta+ \cS^s\bm{\nu}^\cc_\ts).
\EEQ
The only current knowledge of the involved matrix elements lies in (\ref{essv=}).

The \underline{ \!f\! }\!lavor eigenstates  can also be written as single sums over \underline{m}ass eigenstates,
\BEQ\label{nufnum}
&& \nu_{\alpha L}=\sum_{i=1}^6U_{\alpha i}{\nu}_{\it mL}^i,\quad \bm{\nu}_{f L} =U\bm{\nu}_{\it mL} ,  \qquad  
 \bm{\nu}_{\it mL}=(\nu_{\ta L}^1,\nu_{\ta L}^2,\nu_{\ta L}^3,\nu_{\ts R}^{1\cc},\nu_{\ts R}^{2\cc},\nu_{\ts R}^{3\cc})^T,
\EEQ 
with the $3\times 6$ PMNS matrix $U$ having elements
\BEQ
\label{U36}
U_{\alpha,i}=A_{\alpha i}=(U^\DpM\cA)_{\alpha i}, \quad U_{\alpha ,i+3}=S_{\alpha i}=(U^\DpM\cS)_{\alpha i},\quad  
 (UU^\dagger)_{\alpha\beta}=\delta_{\alpha\beta},\quad (\alpha,\beta=1,2,3).
 \EEQ
the latter deriving from (\ref{AApSS=}), while $U^\dagger U\neq \veen_{6\times6}$,
because $U$ represents the 3 active rows of a unitary $6\times6$ matrix which also involves $\cA^s$ and $\cS^s$.
Hence the GIM theorem that $J_{NC}^{\,\mu}$ has the same form on flavor and mass basis,
does not hold\cite{giunti2007fundamentals}.

Inserted in the currents the relations (\ref{nuf2num}) yield
\BEQ
 \label{JccAS}
J^{\mu}_{CC}&=& 2\overline{\bm{\nu}_{mL}}\, U^\dagger\gamma^{\,\mu}\bm{\ell}_{ L}
=2(\overline{\bm{\nu}_{\ta L}}A^\dagger+\overline{\bm{\nu}_{\ts R}^\cc}S^\dagger)\gamma^{\,\mu}\bm{\ell}_{ L}
=2(\overline{\bm{\nu}_{\ta L}}\cA^T
-\bm{\nu}_{\ts R}^T\cC^\dagger\cS^T)U^{\DpM\dagger}\gamma^{\,\mu}\bm{\ell}_{ L}
\nn \hspace{1.4cm} (61a)
\\ \label{JccASdag}
J^{\mu\,\dagger}_{CC}&=& 2\overline{\bm{\ell}_{ L}} \gamma^{\,\mu}
 U\bm{\nu}_{mL} \,\,\, =2\overline{\bm{\ell}_{ L}} \gamma^{\,\mu}(A \bm{\nu}_{\ta L}+S \bm{\nu}_{\ts R}^{c})  \,\,\,\, =
 2\overline{\bm{\ell}_{ L}} \gamma^{\,\mu} 
U^\DpM(\cA \bm{\nu}_{\ta L}+\cS \bm{\nu}_{\ts R}^{c}) ,   \nn \hspace{2.3cm} (61b)
\\ \label{JncAS}
J_{NC}^{\,\mu}& =& \overline{\bm{\nu}_{\ta L}}\gamma^{\,\mu} \cA^T \cA\,\bm{\nu}_{\ta L} 
-\overline{\bm{\nu}_{\ts R}}\gamma^{\,\mu}\cS^T \cS\bm{\nu}_{\ts R} 
 -\bm{\nu}_{\ts R}^T\cC^\dagger\gamma^{\,\mu} \cS^T \! \cA\,\bm{\nu}_{\ta L} 
-\bm{\nu}_{\ta L} ^\dagger \gamma^{\mu\dagger} \cC \cA^T\! \cS \bm{\nu}_{\ts R}^{\dagger T} . \nn \hspace{2.05cm} (61c)
\EEQ

\subsection{Lepton number for sterile neutrinos}

There is an ambiguity in defining the lepton number of the sterile neutrinos.
The lepton number of neutrinos is investigated by making the transformation 
\setcounter{equation}{61}
\BEQ
\bm{\nu}_{aL}\to e^{\ci \rmL_a\phi}\bm{\nu}_{aL},\qquad 
\bm{\nu}_{sR}\to e^{\ci \rmL_s\phi}\bm{\nu}_{sR},\qquad 
\EEQ
This leaves the kinetic terms invariant and for the standard choice $\rmL_s=\rmL_a=1$ also the Dirac mass terms (\ref{LDirac}). 
Only the Majorana mass terms (\refl{LmML}) and (\refl{LmMR}) will vary by factors $e^{\pm 2 \ci \phi}$: they violate lepton number conservation by $\Delta \rmL=\pm 2$.
This approach connects the lepton number $\rmL_a=1$ of active neutrinos also to sterile neutrinos, hence $\rmL_{\bar\nu_s}=-1$ for sterile antineutrinos 
(charge conjugated sterile ones). This assigns lepton number $+1$ to the components $j=1,2,3$ of $\bN_{fL}$ of Eq. (\ref{NfL}) and $\bn_{L}$ of Eq. (\ref{nL=}),
but $-1$ to the components $j=4,5,6$. Then the mixing (\ref{numanums}), or its reverse (\ref{nuf2num}), (\ref{nusnusa}),
enforced by the nonvanishing right handed Majorana mass matrix, makes it impossible to consistently connect a lepton number to the 
particles connected to the mass eigenstates $\nu_\ta^i$ and $\nu_\ts^i$. 

The opposite choice $\rmL_a=1$, $\rmL_s=-1$ circumvents this problem for general models with active and sterile neutrinos.
According to (\ref{NfL}),  (\ref{numanums})  and (\ref{nuf2num}) the lepton number $\rmL_a = 1$ of  $\bm{\nu}_a$ particles is consistent 
with $\rmL_\ta = 1$ of a $\bm{\nu}_\ta$ particle and $\rmL_\ts=-1$ of  $\bm{\nu}_\ts$ particles. This choice is henceforward consistent with (\ref{nusnusa}).
The benefit of this convention is that in pion and neutron decay both channels 
$\pi^-\to \mu+\bar\nu_{\ta L}$, $\pi^-\to \mu+\nu_{\ts R}$, and  
 $n\to p+e+\bar\nu_{\tilde e}$, $n\to p+e+\nu_\ts$, respectively, satisfy lepton number conservation.

The minor price to pay is that now both the Dirac mass term and the Majorana terms violate  lepton number conservation by two units, 
so that the unsolvable problem of lepton number violation remains unsolved. Indeed, as we shall discuss below, 
the Majorana mass terms still allow for neutrinoless double $\beta$ decay,
where a nucleus decays by emitting two electrons (or two positrons) but no (anti)neutrinos.

\section{Applications}

\subsection{Estimates for the Dirac and Majorana masses}
\label{massestimates}

For later use we present the eigenvalues up to third order in $\mu_{1,2,3}$ and $\mub_{1,2,3}$, viz.
\BEQ  \label{eigvls}  
&&
\hspace{-5mm}
 \lambda_1^\pm=\pm \mDe \Big(1-\frac{\mu_2^2}{2\Deltat}+\frac{\mu_3^2}{2\Deltad} \Big) +\frac{\mube}{2}- \frac{\mDe ^2 \mu_1 \mu_2 \mu_3 }{\Deltat\Deltad},
\nn\\ && 
 \hspace{-5mm}
\lambda_2^\pm=\pm \mDt \Big(1 - \frac{\mu_3^2}{2\Deltad}+\frac{\mu_1^2}{2\Deltae } \Big) +\frac{\mubt}{2}-\frac{\mDt ^2 \mu_1 \mu_2 \mu_3 }{\Deltad \Deltae} ,
\\&&
 \hspace{-5mm}
\lambda_3^\pm=\pm \mDd \Big(1- \frac{ \mu_1^2}{2\Deltae }+\frac{ \mu_2^2}{2\Deltat} \Big )+\frac{\mubd}{2}-\frac{\mDd ^2 \mu_1 \mu_2 \mu_3 }{\Deltae \Deltat}   . \nn
\EEQ
Due to Eq.~(\ref{mubmui=})  the last terms cancel,  to make the $m_i$ $=$ $|\lambda_i^\pm|$ pairwise degenerate.
Employing the averages $\mb^2=\frac{1}{3}(\mbe^2+\mbt^2+\mbd^2)$ and $m^2=\frac{1}{3}(m_1^2+m_2^2+m_3^2)$, 
the mass-squared differences $\Delta m_{ij}^2\equiv m_i^2-m_j^2$  become~approximately, 
\BEQ\label{Delta123}
&& \hspace{-1cm}
\Delta_1\equiv \Delta m^2_{23}=\Deltae+\mb^2\Big(\frac{2\mu_1^2}{\Deltae} -\frac{\mu_2^2}{\Deltat} -\frac{\mu_3^2}{\Deltad} \Big), \quad 
\Delta_2\equiv \Delta m^2_{31}=\Deltat +\mb^2\Big(\frac{2\mu_2^2}{\Deltat}  -\frac{\mu_3^2}{\Deltad} -\frac{\mu_1^2}{\Deltae} \Big), \nn
\nn \hspace{4.5mm} (64a,b)
 \\&& \hspace{-1cm}  \Delta_3\equiv \Delta m^2_{12}=\Deltad+\mb^2\Big(\frac{2\mu_3^2}{\Deltad} -\frac{\mu_1^2}{\Deltae} -\frac{\mu_2^2}{\Deltat} \Big) , 
\nn \hspace{9.1cm} (64c)
\EEQ
provided that $\mu_i^2\ll |\bar\Delta_i|$. It holds that $-\Delta_3=\Delta m^2_{\rm sol}=(7.53\pm0.18)\,10^{-5} \eV^2$\cite{tanabashi2018review}.
Normal ordering $m_1<m_2< m_3$ is connected to $-\Delta_1=\Delta m^2_{\rm atm}= (2.44\pm 0.06)\,10^{-3} \eV^2$\cite{tanabashi2018review} and 
$\Delta_2=-\Delta_1-\Delta_3>0$,
while inverse ordering $m_3<m_1<m_2$ leads to $\Delta_1=\Delta m^2_{\rm atm}$ and $\Delta_2<0$.
Cluster lensing puts forward a value $\mb\sim 1.5 -1.9$ eV for the absolute scale of the neutrino 
masses\cite{nieuwenhuizen2009non,nieuwenhuizen2013observations,nieuwenhuizen2016dirac,nieuwenhuizen2017subjecting}.

With $\varepsilon_{ijk}$ the Levi-Civita symbol, there hold the exact relations 
\setcounter{equation}{64}
\BEQ
\bar m_i^2=\bar m^2+\frac{1}{3}\sum_{j,k=1}^3\varepsilon_{ijk}\bar\Delta_k,\qquad m_i^2=m^2+\frac{1}{3}\sum_{j,k=1}^3\varepsilon_{ijk}\Delta_k.
\EEQ

Let us investigate Eq. (\ref{Delta123}) for normal ordering. 
The effects of the Majorana masses are anticipated to occur at the level of $\Delta m^2_{\rm sol}$.
We fix $\mbt$ and express $\mb_{1,3}$ in $d_{1,3}$ as
\BEQ
\mbe^2=\mbt^2-d_3\Delta m^2_{\rm sol},\qquad 
\mbd^2=\mbt^2+\Delta m^2_{\rm atm}-d_1\Delta m^2_{\rm sol},\qquad 
\EEQ
so that 
\BEQ
-\Deltae=\Delta m^2_{\rm atm}-d_1\Delta m^2_{\rm sol},\qquad -\Deltad=d_3\Delta m^2_{\rm sol}.
\EEQ  
With $\mb\approx \mbt$ we set also
\BEQ \label{mu3Theta3} \hspace{-5mm}
\Theta_3=\frac{\mu_3\mb}{d_3\Delta m^2_{\rm sol}} 
=\frac{ \Delta m^2_{\rm atm}}{\Delta m^2_{\rm sol}} \, \frac{u_3}{d_3},
\qquad 
\mu_3=d_3\Theta_3\frac{\Delta m^2_{\rm sol}}{\mb} ,
\qquad   u_3=d_3\Theta_3\frac{\Delta m^2_{\rm sol}}{\Delta m^2_{\rm atm}} .
\EEQ
With $d_{1,3}$ and $\Theta_3$ fixed we can determine $\mu_{1,2}$ or, equivalently,  $u_{1,2}$.  
Imposing $\Delta m^2_{\rm sol}\ll \Delta m^2_{\rm atm} \lll \mb^2$, we deduce from (\refl{Delta123}c) 
and from the difference of (\refl{Delta123}a) and (\refl{Delta123}b) that
\BEQ &&
u_1^2=\frac{2d_1-5(1-d_3)}{12d_3}+\Theta_3^2,\qquad \mu_1=\sqrt{| \Deltat \Deltad|}\frac{u_1}{\mb_1}, 
\nn\\&&
u_2^2=\frac{2d_1+7(1-d_3)}{12d_3}-\Theta_3^2,\qquad \mu_2=\sqrt{|\Deltad \Deltae |}\frac{u_2}{\mb_2},
\\&&
u^2=u_2^2-u_1^2-u_3^2=\frac{1-d_3}{d_3}-2\Theta_3^2-d_3^2\Theta_3^2\Big(\frac{\Delta m^2_{\rm sol}}{\Delta m^2_{\rm atm}} \Big)^2. \nn
\EEQ 
These are expressions of order unity and exact to leading order in $d_1$, $d_3-1$ and $\Theta_3$. 
Hence the typical scale is $\mu_{1,2}\sim \sqrt{\Delta m^2_{\rm atm}\Delta m^2_{\rm sol}}/\mb=4.3\,10^{-4} \eV^2/\mb$ and $\mu_3\sim 8\, 10^{-5}\eV^2/\mb$.

In the eigenvectors (\ref{eigvcs}) the coefficients 
\BEQ &&
\frac{\bar \mu_{1}}{4\mb}\approx-\frac{\bar \mu_{2}}{4\mb}\approx -u_1u_2u_3\frac{\Delta m^2_{\rm atm}}{2\mb^2},\qquad 
\frac{\bar \mu_{3}}{4\mb}\approx u_1u_2u_3\frac{\Delta m^2_{\rm sol}}{2\mb^2} ,
\EEQ
are typically rather small. Hence the mixing matrix $\cS$ will essentially involve elements $\pm\Theta_{1,2,3}$
\BEQ \label{Theta12=}
 \Theta_{1,2}=\frac{\mb \mu_{1,2}}{|\overline{\Delta}_{1,2}|} 
\approx u_{1,2} \sqrt{\frac{\Delta m^2_{\rm sol}}{\Delta m^2_{\rm atm}}}=0.18u_{1,2},\qquad
\Theta_3\approx 32\frac{u_3}{d_3}.
\EEQ
with $\Theta_3$ introduced in (\ref{mu3Theta3}).
If one of the $\Theta_i$ dominates but is still small, it can be seen as a mixing angle.
In particular $\Theta_3\sim$ 0.1  is possible, which is relevant for the ANITA events to be discussed below.

\subsection{Neutrino oscillations in vacuum}
\label{nu-osc}

We consider neutrino oscillations in the plane wave approximation. See Ref. \cite{akhmedov2009paradoxes} for an excellent discussion of its merits.
In the notation (\ref{nufnum}), (\ref{U36}) an initially pure active state vector reads
\BEQ
|\nu_{\alpha L}(0){\rangle }=\sum_{i=1}^6 U^\ast_{\alpha i}|\nu_{\it mL}^i\rangle.
\EEQ
It evolves after time $t$  into
\BEQ\hspace{-7mm}
|\nu_{\alpha L}(t)\rangle=\sum_{i=1}^6 U^\ast_{\alpha i}e^{-\ci\phi_i}|\nu_{mL}^i\rangle ,
\EEQ
with the Lorentz invariant phase at  $|\vr|\approx ct$ given by
\BEQ
\phi_i=\frac{E_it-\vp\ccdot\vr}{\hbar}\approx \Big(\sqrt{1+\frac{m_i^2c^2}{p^2}}-1\Big)\frac{pct}{\hbar}\approx \frac{m_i^2c^3t}{2\hbar p}.
\EEQ
This result can be motivated for a wave packet \cite{akhmedov2009paradoxes}.
The amplitude for transition into active state $\beta$ is 
\BEQ 
\langle \nu_{\beta L} |\nu_{\alpha L}(t)\rangle &=&
\sum_{i=1}^6 U _{\beta i}U^\ast_{\alpha i}e^{-\ci\phi_i} 
= \sum_{i,k,l=1}^3 U^D _{\beta k}U^{D\ast}_{\alpha l}(\cA_{ki}\cA_{li}+\cS_{ki}\cS_{li} )e^{\ci(\eta_k-\eta_l)-\ci\phi_i}.  
\EEQ
where we used that $\phi_{2i-1}=\phi_{2i}$. The transition probability after time $t$ may be expressed as two terms, 
\BEQ
P_{\nu_\alpha\to\nu_\beta}^\DpM=|\langle \nu_{\beta L} |\nu_{\alpha L}(t)\rangle |^2
=P_{\nu_\alpha\to\nu_\beta}^D+P_{\nu_\alpha\to\nu_\beta}^M ,
\EEQ
where the first one
\BEQ
P_{\nu_\alpha\to\nu_\beta}^D=\delta_{\alpha\beta}- \sum_{i,j=1}^3U^D _{\beta i}U^{D\ast}_{\alpha i}U^{D\ast}_{\beta j}U^D_{\alpha j} (1-e^{\ci\phi_j-\ci\phi_i} ) ,
\EEQ
 represents the standard ``Dirac'' result and where the Majorana masses add the ``Majorana'' expression
\BEQ
\hspace{-2mm} \label{PabM=}
P_{\nu_\alpha\to\nu_\beta}^M &=&\sum_{i,j,k,l,m,n=1}^3
U^D _{\beta k}U^{D\ast}_{\alpha l}U^{D\ast}_{\beta m}U^D_{\alpha n} 
e^{\ci(\eta_k-\eta_l-\eta_m+\eta_n)}(e^{\ci\phi_j-\ci\phi_i} -1)   \nn
\\&& \hspace{18mm}
\times\hspace{2mm}[(\cA_{ki}\cA_{li}+\cS_{ki}\cS_{li})(\cA_{mj}\cA_{nj}+\cS_{mj}\cS_{nj})-\delta_{ki}\delta_{li}\delta_{mj}\delta_{nj}]  .   
\EEQ
 The fact that $\sum_{\beta=1}^3P_{\nu_\alpha\to\nu_\beta}^M\leq0$ reflects oscillation into sterile states.

While the Majorana phases $\eta_i$ cancel in $P_{\nu_\alpha\to\nu_\beta}^D$ as usual, they remain present in $P_{\nu_\alpha\to\nu_\beta}^M$.
This occurs in the \DMnSM \,  because they are upgraded to physical Dirac phases, see footnote \ref{fntMaj}.
$P_{\nu_\alpha\to\nu_\beta}^\DpM$ has the schematic $\delta$-dependence $1+\cos\delta+\sin\delta+\cos2\delta+\sin2\delta$,
but the $\cos2\delta$, $\sin2\delta$ terms are turned into $\cos\delta$, $\sin\delta$ terms by taking $\eta_3\to\eta_3'=\delta+\eta_3$.
This is equivalent to replacing the $U^D$ of (\ref{Udecomp}) by
\BEQ\hspace{-5mm}
\tilde U^D&=&U_1U_2U_3\times{\diag}(1,1,e^{\ci\delta})
=\begin{pmatrix}  
c_2 c_3 & c_2 s_3 & s_2 \\
 - c_1 s_3 -s_1 s_2c_3e^{\ci\delta}  &  c_1 c_3 - s_1 s_2 s_3 e^{\ci\delta}    & s_1 c_2e^{\ci\delta} \\
\,\,\  s_1 s_3 -c_1  s_2c_3 e^{\ci\delta}   & -s_1 c_3-    c_1 s_2 s_3 e^{\ci\delta} & c_1 c_2 e^{\ci\delta}
   \end{pmatrix} , 
\EEQ
wherein $\delta$ enters only in the schematic form $1+e^{\ci\delta}$.
 Absolute Majorana phases have no physical meaning; indeed,  the expression  (\ref{PabM=}) involves them only through their differences.
 The CP violation effect,
\BEQ\label{DeltaPabCP}
\Delta P_{\nu_\alpha\leftrightarrow\nu_\beta}^{CP}
=P_{\nu_\alpha\to\nu_\beta}^\DpM-P_{\bar\nu_\alpha\to\bar\nu_\beta}^\DpM
=P_{\nu_\alpha\to\nu_\beta}^\DpM-P_{\nu_\beta\to\nu_\alpha}^\DpM ,
\EEQ
can be read off from the above by switching $\alpha\leftrightarrow\beta$. 
The terms with $\sin (2\eta_i-\eta_j-\eta_k)$, $\cos(2\eta_i-\eta_j-\eta_k)$  with $j=k$ and $j\neq k$ from (\ref{PabM=}) cancel, 
leaving the dependence on $\delta$, the $\eta_i$ and $t$ of the  form
\BEQ \label{DeltaPabCPv}
\Delta P_{\nu_\alpha\leftrightarrow\nu_\beta}^{CP}(t)=\sum_{k=1}^3
\Big[
d_k\sin\delta+\sum_{i\neq j=1}^3d_{ijk}\sin(\delta+\eta_i-\eta_j)+\sum_{i>j=1}^3c_{ijk}\sin(\eta_i-\eta_j)\Big]\sin\frac{\Delta_kt}{2q} .
\EEQ
Choosing $\eta_1=0$ this vanishes only for the trivial values $\delta$, $\eta_{2}$, $\eta_{,3}$ equal to $0$ or $\pi$.
It confirms that in the \DMnSM \, two of the Majorana phases $\eta_{i}$ of the \SnuM \, are physical Dirac phases$^{\ref{fntMaj}}$.

\subsection{Neutrino oscillations in matter}

Relativistic neutrinos have energy $ E_i=(q^2+m_i^2)^{1/2}$ $\approx q+{m_i^2}{/2q}$.
Neutrino oscillations in vacuum are ruled by  the Hamiltonian $H_0=E$, which reads on the flavor basis 
\BEQ\hspace{-5mm}
(H_0)_{\alpha\beta}
\approx \sum_{i=1}^6U_{\alpha i}\big(q+\frac{m_i^2}{2q}\big)|\nu_{mL}^i\rangle \langle\nu_{mL}^i | U^\dagger_{i\beta} .
\EEQ
The $q$ term leads to $q\delta_{\alpha\beta}$ and can be omitted, as it plays no role for the eigenfunctions.
For propagation in matter one adds the matter potential. The charged and neutral currents induce the scalar potentials
\BEQ
V_{CC}=\sqrt{2}G_Fn_e,\quad V_{NC}=-\frac{1}{2}\sqrt{2}G_Fn_n,
\EEQ
 involving the electron number density $n_e=n_{e^-}-n_{e^+}$
and the neutron number density $n_n$, and yielding
\BEQ
V_e=V_{CC}+V_{NC},\quad V_\mu=V_\tau=V_{NC},\quad 
\EEQ
The potential of the active neutrinos is diagonal on the flavor basis, while the sterile ones 
do not sense any. This results in the total matter potential on the flavor basis
\BEQ \hspace{-5mm}
V_m={\rm diag}(V_e,V_\mu,V_\tau,0,0,0) .
\EEQ

From Eq. (\ref{MDpM=}) and its real, symmetric nature it follows that 
\BEQ
M^{\DpM}=U_{LR}^\ast\, \cM \, U_{LR}^\dagger,\qquad
M^{\DpM\,\dagger}=U_{LR} \cM \, U_{LR}^T.
\EEQ
Hence the $m_i^2$, the eigenvalues of $\cM^2$, arise from 
\BEQ
M^{\DpM\,\dagger}\,M^{\DpM}=U_{LR}\ \cM^2 \, U_{LR}^\dagger .
\EEQ
The total matter Hamiltonian therefore reads on the flavor basis
\BEQ\label{Hm=}
H_m=\frac{1}{2q}M^{\DpM\,\dagger}\,M^{\DpM}+V_m .
\EEQ
Let us set $\cV_m=2qU_{LR}^\dagger V_mU_{LR}=(\cV_m^a,\cV_m^s)$ with $\cV_m^s={\rm diag}(0,0,0)$ and 
\BEQ
\cV_m^a=2q \, U_L^\dagger\,V_m^a\, U_L  ,\qquad V_m^a={\rm diag}(V_e,V_\mu,V_\tau). \qquad 
\EEQ
The factor $D_L'$ in the  decomposition (\ref{Udecomp}) for $U_L$, also drops out from $\cV_m^a$ since $V_m^a$ is diagonal,
hence it can be totally omitted.
Eq. (\ref{Hm=}) can be expressed as
\BEQ\label{cHm=} &&
H_m=U_{LR}\, \cH_m \, U_{LR} ^\dagger, 
\qquad 
 \cH_m= \frac{\cM^2 +\cV_m}{2q} =\frac{1}{2q}\begin{pmatrix} M^{d\, 2}+\cV_m^a & M^d M^N \\ M^NM^d & M^{d\,2}+M^{N\,2} \end{pmatrix} . 
\EEQ
First solving the eigenmodes of $\cH_m$ and then going to the flavor basis allows to evaluate the effects of oscillations on the active neutrinos 
without having knowledge of the undetermined matrix $U_R$. The eigenfunctions do not alter upon subtracting $(\mb^2/2q)\,\veen_{6\times 6}$ 
from $\cH_m$, after which all elements are small.

Inside matter the Diracian properties are lost, there are just 6 Majorana states with different masses.
While the neutral current potential $V_{NC}$ can be omitted in the limit $M^N\to0$, this is not allowed in general.
In matter one has real potentials $V_\alpha$, $\alpha=e,\mu,\tau$. Due to $V_{CC}$  the matrix $\cV_m^a$ is complex hermitian.
The hermitian $ \cH_m$  has 6 different positive eigenvalues but complex valued eigenmodes.

Inside the Sun the neutrino transport is dominated by a mostly electron-neutrino mode\cite{smirnov2016solar}; 
in the 3+3 model this is represented by two nearly degenerate, nearly maximally mixed Majorana modes. 
The resonance condition in the standard solar model now splits up as a condition for each of them. 

The so-called solar abundance problem stems from the inconsistency between the standard solar model parameterized 
by the best description of the photosphere and the one parameterized to optimize agreement with
helioseismic data sensitive to interior composition\cite{basu2009fresh}.
The biggest deviations in the solar composition are of relative order $1\%$ and occur at $\sim0.7R_\odot$.
Our modified resonance conditions offer hope for an improved description of the data.

\subsection{Pion decay}
\label{piondecay}

One of the simplest elementary particle reactions is  
\BEQ\label{exitpimin}
\pi^-\to W^-\to \mu+\bar\nu_{\mu R}. \qquad 
\EEQ
It describes a negatively charged $\pi^-$ particle, $\pi^-=(d\bar u)$, consisting of a down quark (charge $-{e}/{3}$) and
an anti-up quark (charge $-{2e}/{3}$), decaying into a $W^-$ boson (charge $-e$), which in its turn decays into a muon (charge $-e$)
and an muon-antineutrino (charge 0). Related reactions are
$\ \pi^-\to e+\bar\nu_e$, $ \pi^+\to \mu^++\nu_\mu$ and $\pi^+\to e^++\nu_e $.
In the \DMnSM \,  the current $J^{\mu\,\dagger}_{CC} =2\overline{\bm{\ell}_{ L}} \gamma^{\,\mu}U\bm{\nu}_{mL}=
2\overline{\bm{\ell}_{ L}} \gamma^{\,\mu}(A \bm{\nu}_{\ta L}+S \bm{\nu}_{\ts R}^{c})$ 
replaces Eq. (\ref{exitpimin}) by decays with any of the 6 mass eigenstates $(\overline{\nu}_{m}^{\, i})_R$ emitted.
They can be grouped as 
\BEQ\label{exitpiminsa}
\pi^-\to W^-\to \mu+\bar\nu_{\ta R}^i , \quad (i=1,2,3), \qquad
\pi^-\to W^-\to  \mu+\nu_{\ts R}^{i-3}  \quad (i=4,5,6) .
\EEQ
The $\nu_{\ta L}^i$ and the charge conjugated $\nu_{\ts R}^{i\cc}$ fields have identical chiral structure 
(up to a phase factor, see Ref. \cite{giunti2007fundamentals}, eqs (2.139) vs. (2.356)), 
differing only by their creation and annihilation operators.
Hence all decay channels involve the standard chiral factors, and a new factor, the sum over final neutrino states,
$\sum_{i=1}^6  |U_{\mu i}|^2=\sum_{i=1}^3  |(U^DD^M\cA)_{\mu i}|^2+\sum_{i=1}^3 |(U^DD^M\cS)_{\mu i}|^2 $. It equals 
\BEQ \label{ASmumu}
\hspace{-4mm}
\big(UU^\dagger\big)_{\alpha\beta}=
\big(U^DD^M(\cA\cA^T+\cS\cS^T)D^{M\dagger}U^{D\dagger}\big)_{\alpha\beta}
=\delta_{\alpha\beta}, 
\EEQ
 for $\alpha=\beta=\mu$. (In this equality we employed Eq. (\ref{AApSS=})).
So charged pions decay in the \DMnSM \, at the same rate as in the SM.
Neutral pion decay does not involve neutrinos, so it is also not modified.

\subsection{Neutron decay}

A neutron $n=(ddu)$ consists of 2 down quarks and one up quark, and a proton $p=(duu)$ of 1 down and 2 up quarks.
Neutron decay $n\to p+e+\bar\nu_e$ involves a transition from a down quark to an up quark producing a virtual $W^-$ boson,
which decays into an electron and an electron antineutrino. 
As coded in the charged current $J^{\mu\,\dagger}_{CC}$ of  (\ref{Jcc=}), it occurs in the DM$\nu$SM in two channels, $n\to p+e+\bar\nu_\ta$ and $n\to p+e+\nu_\ts$. 
Both decay channels involve the standard chiral factors, and a new factor, the sum over final neutrino states 
$\sum_i  |(U^DD^M\cA)_{ei}|^2+\sum_i |(U^DD^M\cS)_{ei}|^2 $. 
This is the $\alpha=\beta=e$ element of Eq.  (\ref{ASmumu}), so it is equal to unity.
Hence the neutron lifetime  in the DM$\nu$SM stems with the one in the \SnuM.

The main decay channel is $n\to p+e+\bar\nu_\ta$. 
With our convention $L_{\nu_\ts}=-1$, also  the channel  $n\to p+e+\nu_\ts$ conserves the lepton number.
The latter occurs at a slower rate due to the small term $\cS\cS^T$ in  (\ref{ASmumu}).
We could not convince ourselves that it would be ineffective in beam experiments
and hence be capable to explain the neutron decay anomaly between beam and bottle measurements\cite{wietfeldt2011colloquium}.

\subsection{Muon decay}

With the neutron decay going into two channels,
muon decay  $\mu\to e+\bar\nu_{eR}+\nu_{\mu L}$ goes into four,
\BEQ &&
\mu\to e+\bar\nu_{\ta R}+\nu_{\ta L},\quad 
\mu\to e+\bar\nu_{\ta R}+\bar \nu_{\ts L} ,\quad
\mu\to e+\nu_{\ts R }+\nu_{\ta L } ,\quad 
 \mu\to e+\nu_{\ts R}+\bar \nu_{\ts L } ,
\EEQ
with rates of leading schematic order $1$, $\cS\cS^T$, $\cS\cS^T$ and $(\cS\cS^T)^2$, respectively,
adding up  to the SM result.

\subsection{Neutrinoless double $\beta$-decay}

In a simultaneous double neutron decay (double $\beta$-decay) the emission of two electrons involves the schematic neutrino terms
${\nu}_{\ta L}^{i\,2}+{\nu}_{\ts R}^{i\cc\,2}+ {\nu}_{\ta L}^i {\nu}_{\ts R}^{i\cc}$,
corresponding to the emission of two mostly active antineutrinos, two mostly sterile neutrinos, or one of each. 
With $L_{\nu_\ta}=1$ and $L_{\nu_\ts}=-1$, all three channels conserve the lepton number.

In the standard neutrino model also neutrinoless double $\beta$-decay is possible. 
Then only the ${\nu}_{\ta L}^{i\,2}$ term occurs, subject to the Majorana condition ${\nu}_{\ta L}^{i\,\cc}={\nu}_{\ta L}^i$;
with $\cA_{ij}\to\delta_{ij}$ and $\cS_{ij}\to 0$,  it yields an amplitude proportional to 
$m_{ee}=\sum_{i=1}^3U_{ei}^{D\,2}e^{2\ci \eta_i}m_i$ for small $m_i$.
The GERDA search puts a bound $|m_{ee}| \le 0.15-0.33$ eV \cite{gerda2017background}.
Does this rule out the \DMnSM \, for $m\sim 2$ eV? Not, as we show now.

 In our situation with Diracian neutrinos  neither ${\nu}_{\ta L}^{i\,2}$ nor ${\nu}_{\ts R}^{i\cc\,2}$ contributes,
but neutrinoless double-$\beta$ decay does arise from the ${\nu}_{\ta L}{\nu}_{\ts R}^\cc$ terms. All spinor terms are again as in the \SnuM.
The only change occurs in the effective mass, which now reads 
\BEQ \hspace{-1mm} \label{mee=}
m_{ee}=[A \underline{m} S^T+S\underline{m}A^T]_{ee} 
=[U^DD^M(\cA \underline{m} \cS^T+\cS \underline{m}\cA^T) D^M U^{D\,T}]_{ee},\quad \underline{m}={\rm diag}(m_1,m_2,m_3).
\EEQ 
It involves cancellations, since $\cS$ is nearly asymmetric while $\cA$ and $\underline{m}/m$ are close to the identity matrix.
But the cancellations are maximal.  From the definitions (\ref{cAcSdef}) we can go back to the six eigenvectors $\ve^{(i)}$ 
of Eqs. (\ref{cMdecomp}) and (\ref{e2im12i}). Recalling that $\lambda_{2i}=m_i$ and $\lambda_{2i-1}=-m_i$, ($i=1,2,3$), it follows that
\BEQ 
(\cA \underline{m}\cS^T+\cS \underline{m}\cA^T)_{jk}&=&
\sum_{i=1}^6\lambda_i e^{i}_je^{i}_k=\cM_{jk} .
\EEQ
From (\ref{cMmatrix}) it is seen that $\cM_{jk}=0$ for $j,k=1,2,3$, because we neglected the left handed Majorana mass matrix  $M_L^M$.
In general $m_{ee}=M^\DpM_{ee}=(M^M_{L})_{ee}$ \cite{gonzalez2008phenomenology}.
Hence $m_{ee}=0$ for this leading order diagram.

Nevertheless, neutrinoless double $\beta$--decay,  involving lepton number violation $\Delta\, \rmL=2$,
is not forbidden in the \DMnSM. It occurs in the $m_i^3/q^2$ correction to $m_i$ in (\ref{mee=}) stemming from the internal propagator 
$m_i/(q^2-m_i^2)=m_i/q^2+m_i^3/q^4+\cdots$. But the suppression factor $m_i^2/q^2$ makes its measurement impractical for realistic $q\sim\MeV-\GeV$.
Loop effects may fare better, but are also tiny. If a finite $m_{ee}$ is established, it points at new high-energy physics.

In conclusion, the non-detection of neutrinoless double $\beta$-decay is compatible with the \DMnSM.

\subsection{Small twin-oscillation}

If the degeneracy of the solar twin modes is slightly lifted by non-cancellation of the last two terms in each line of Eq. (\ref{eigvls}), one gets
\BEQ
\Delta \mDe ^2\equiv (\lambda_1^+)^2-(\lambda_1^-)^ 2\approx
\mD \mube-\frac{2\mD ^3\mu_1 \mu_2 \mu_3  }{\Deltat\Deltad}  .
\EEQ
From the standard solar model we know that oscillations should not occur underway to Earth\cite{VinyolesNewGeneration2017}, 
so that $|\Delta \mDe ^2|\lesssim 10^{-12}\,\eV^2$.
For the supernova SN1987A at distance of $51.4$ kpc the absence of twin-oscillations even implies that  $|\Delta \mDe ^2|\lesssim 10^{-22}\,\eV^2$;
the alternative is that twin-oscillation did take place, and only half of the emitted neutrinos arriving here on Earth were active and could be detected.
The implied doubling of power emitted in neutrinos then requires  an adjustment of the SN1987A explosion model. 

The similarly defined $\Delta \mDt ^2$ and $\Delta \mDd ^2$ may be larger. 
Either of them may describe the MiniBooNE anomaly \cite{aguilar2018significant} (disputed by MINOS \cite{adamson2019search} 
and still debated \cite{boser2019status})
with $\Delta m^2\approx 0.04\, \eV^2$. But a more elegant approach hereto is to keep the 3 Diracian neutrinos and add a fourth sterile one.

\subsection{Sterile neutrino creation in the early Universe}

The creation of sterile neutrinos in cosmology is an important process based on loss of coherence in oscillation processes.
It is well studied, see e.g. \cite{lesgourgues2013neutrino}, and is important when sterile neutrinos are to make up half of the cluster dark 
matter\cite{nieuwenhuizen2009non,nieuwenhuizen2013observations,nieuwenhuizen2016dirac,nieuwenhuizen2017subjecting}.

The charged currents (\ref{JccASdag}a,b) allow the creation of sterile neutrinos out of active ones via $e^++\nu_e\to W^+\to e^++\bar\nu_s$, and creation of 
sterile antineutrinos out of active ones in the process $e+\bar\nu_e\to W^-\to e+\nu_s$, that is, by four-Fermi processes with virtual $W^\pm$ exchange.
The first term in Eq. (\ref{JccAS}c) describes interaction of active neutrinos with $Z$, the second of sterile ones, and the last two the exchange of active versus
sterile neutrinos, and vice versa. In particular the creation of sterile neutrinos out of active ones is possible in two channels via the four-Fermi process 
$\nu_\alpha+\bar\nu_\alpha\to\nu_s+\bar\nu_s$ with the exchange of a virtual $Z$ boson. All these processes conserve lepton number.
As is seen from the sterile component in the flavor eigenstate (\ref{nuf2num}) or from the charged and neutral currents (\ref{JccASdag}),
to achieve the sterile neutrino creation a finite matrix $\cS$ is needed.
Hence it does not occur in the standard Dirac limit where both Majorana mass matrices $M_L^M$ and $M^M_R$ vanish.

\subsection{Muon $g-2$ anomaly}

The gyromagnetic factor of the muon is $g_\mu=2(1+a_\mu)$.
Dirac theory yields $g_\mu=2$ and $a_\mu$ is the anomaly due to quantum effects. The leading term is Schwinger's famous result,
\BEQ
a_\mu=\frac{\alpha}{2\pi}+\cdots = 0.00116+\cdots.
\EEQ
$a_\mu$ is known up to its 9th digit, but there is a $\sim 3.5\sigma$ discrepancy between measurement and prediction,
$a_\mu^{\rm exp}-a_\mu^{SM}= 288(63)(43)\,10^{-11}$ \cite{tanabashi2018review}, where the first error is statistical and the second systematic.

Our interest lies in the contribution of neutrinos, which occurs in a simple triangle diagram with virtual $W$ bosons.
Ref. \cite{abdallah2012muon} presents the result for an arbitrary number of sterile neutrinos. For neutrino masses well below $ M_W$ it reads
\BEQ\label{amunu}
a_\mu^\nu&=&(a_\mu^\nu)^{\rm SM}
\sum_{i=1}^{3+N_s} U_{\mu i}U_{\mu i}^\ast
=(a_\mu^\nu)^{\rm SM}(UU^\dagger)_{\mu\mu}
=(a_\mu^\nu)^{\rm SM}
=\frac{G_F}{\sqt}\frac{5m_\mu^2}{12\pi^2} =389\, 10^{-11}.
\EEQ
The unitarity relation (\ref{U36}), viz. $(UU^\dagger)_{\alpha\beta}=\delta_{\alpha\beta}$, is valid even beyond our $N_s=3$ 
case\cite{giunti2007fundamentals}. So the \DMnSM \,  reproduces the one-loop outcome of the SM,
as well as the dominant two-loop electroweak contributions of Ref. \cite{kukhto1992dominant}.

\subsection{ANITA detection of UHE cosmic neutrino events}

Scattering of ultra high energy (UHE) cosmic rays on cosmic microwave background photons puts the GZK 
limit on their maximal energy\cite{greisen1966end,Zatsepin:1966jv} 
and acts as a source for EeV ($10^{18}$ eV) (anti)neutrinos via the creation and decay of charged pions\cite{Beresinsky:CosmicRays}, 
as considered in section \ref{piondecay}.

The Antarctic Impulsive Transient Antenna (ANITA) is a balloon experiment at the South Pole 
that detects the radio pulses emitted when UHE cosmic neutrinos interact with the Antarctic ice sheet. In a set of $\gtrsim 30$ cosmic ray events,
ANITA has discovered an {\it upward} going event with energy $E\sim0.6$  EeV \cite{gorham2016characteristics} and one with
$\sim 0.56$ EeV \cite{gorham2018observation}. Both are consistent with the cascade caused by a $\tau$ lepton created beneath the ice surface\footnote{See 
\cite{shoemaker2019reflections} for a possible explanation due to sub-surface reflection in the ice for a downward event.}.
But the SM connects a relatively large neutrino-nucleon cross section $\sigma\sim G_F^2m_NE$ to an UHE neutrino \cite{connolly2011calculation}, so that
the probability for it to  traverse a large path $L$ through the Earth to reach Antartica is $P_T\sim \exp({-n \, \sigma L})$ is small,
where  $n$ is the effective density of nuclei. The numbers are $P_T\sim 4\,10^{-6}$ and $2\,10^{-8}$, 
respectively \cite{gorham2016characteristics,gorham2018observation}.
In Ref. \cite{cherry2018sterile}  it is pointed out that a sterile neutrino with a smaller cross section, viz. $\sigma\to\sigma \sin^2\Theta$,
with $\Theta$ the mixing angle with respect to active neutrinos, may be involved.
To explain the events and relate them to the detections at IceCube, AUGER  and Super-Kamiokande, these authors fix $\Theta$ at $0.1$  \cite{cherry2018sterile}.

For typical models a sterile neutrino with such a large mixing angle  should have been discovered already.
 In the \DMnSM \, the situation is different, however. It contains the reaction $\nu_\ts\to \tau+W^+$, where the $W^+$ is quickly lost locally 
 but the $\tau$ escapes and decays while creating a shower.
In the approximation where one mixing angle dominates, an emitted electron neutrino will have components 
 of strength $\sin^2\Theta$ on mostly sterile states. Inside the Earth  they are scattered less, and, given that they enter the other side of the Earth,
  can be measured in the $\tau$-flavor mode with modified probability 
  $\tilde P_T\sim\sin^2\Theta\,\exp(-n \, \sigma L \sin^2\Theta)$ and modified flux   $\Phi_{\rm sterile}/\Phi_{\rm active}\sim\sin^4\Theta\,\exp(-n \, \sigma L \sin^2\Theta)$. 
The estimates of section \ref{massestimates} show that the value $\Theta=0.1$ is reasonable for the component $\Theta_3$
of the mixing matrix $\cS$, see in particular the expression for $\Theta_3$ in Eq. (\refl{Theta12=}).
 Hence the  \DMnSM \, supports the sterile-neutrino interpretation of ANITA events. 
 
 For determining the \DMnSM \, parameters, it seems worth to include UHE data.

\section{Summary and outlook}

Since the maximal mixing of pseudo Dirac neutrinos runs into observational problems, neutrino mass is often supposed to stem from a high-energy 
sector beyond the standard model (BSM), for instance by the seesaw mechanism\cite{giunti2007fundamentals,lesgourgues2013neutrino}. 
We show that  the mixing effects can be suppressed in  the Dirac-Majorana neutrino standard model (\DMnSM \,),
the minimal extension of the SM with 3 sterile neutrinos (3+3 model) with both a Dirac and a right handed Majorana mass matrix.
Indeed, to have the 6 physical masses condense in 3 degenerate pairs poses 3 conditions, which leaves 3 Dirac and 3 Majorana masses free.
In this Diracian limit the neutrino mass eigenstates act as Dirac particles like the other fermions in the SM. 
There is no change in the pion, neutron and muon decay, nor on the muon $g-2$ problem.
Compared to the general case, less mixing occurs since members of the same Dirac pair undergo no mutual oscillation.
For small Majorana masses the left handed mass eigenstate is still mostly active, and the right handed one still mostly sterile. 
A flavor eigenstate has a component on mass eigenstates with a mostly sterile character. 
With mixing angles up to 0.2 -- 0.3, this allows to explain the ANITA ultra high energy events.
Hence  for determining the \DMnSM \, parameters, it is natural to include UHE data.

In the Diracian limit the model keeps some of its Majorana properties.
Neutrino oscillations in matter involve the usual 6 nondegenerate Majorana states. Lepton number is not conserved.
Neutrinoless double-$\beta$ decay remains possible, be it at an impractically small rate.
Sterile neutrino generation in the early cosmos is possible at temperatures in the few MeV range.

It is interesting to investigate whether processes involving the neutral current can further test the model.
They are relevant e.g. in nonresonant sterile neutrino production in the early universe.

By connecting lepton number $\rmL=1$ to (mostly) active neutrinos but $\rmL=-1$ to (mostly) sterile neutrinos, 
neutron decay and double $\beta$-decay conserve the lepton number, while lepton number violation is restricted to feeble
neutrinoless double-$\beta$ decay. Should that be observed, it would invalidate our assumption of negligible left handed Majorana mass matrix,
and prove the presence of BSM physics in the high energy sector.

The SM has 19 parameters while 6 neutrino parameters are established and 2 anticipated\voetnoot{The SM  in Eq. (\ref{DMnSM}) has 3 gauge coupling constants; 
2 Higgs self couplings; 6 quark masses;  3 charged lepton masses; 3 strong mixing angles and a strong Dirac phase. Parameter 19 is 
the strong CP angle.  The established neutrino parameters are 2 mass-squared differences, 3 weak mixing angles and,
to some extent, the weak Dirac phase\cite{tanabashi2018review}.
The 2 weak Majorana phases can in the \SnuM \, only be measured via neutrinoless double $\beta$ decay, but in the \DMnSM \, in many ways.}. 
The \DMnSM \, adds 3 further Majorana masses. 
In the limit where they vanish, the sterile partners decouple and the standard neutrino model emerges. 
The extra Majorana masses and Dirac role of the ``Majorana'' phases may alleviate some of the tensions in solar, reactor and other neutrino problems.

From a philosophical point of view, we do not consider the values of the Dirac and Majorana masses and phases as problematic properties 
in urgent need of an explanation, but rather as further mysteries of the standard model.

\section*{Acknowledgements}
The author is grateful for inspiring lectures by and discussion with his teachers Martinus `Tini' Veltman and Gerardus `Gerard' 't Hooft.

\section*{Appendix A: gamma and charge conjugation matrices}
\label{AppA}

The four $4\times4$ anticommuting $\gamma$-matrices were introduced by Dirac. They play a role in the description of, e.g., 
the chiral left handed and right handed electron and positron.

In the convention of Giunti and Kim\cite{giunti2007fundamentals}
the Lorentz indices $\mu=0,1,2,3$ label the coordinates $x^{\,\mu}=(ct,x,y,z)$. 
The anticommutation relations read for any representation of the $\gamma$ matrices,
\BEQ \hspace{-7mm}
\{\gamma^{\,\mu},\gamma^\nu\}=2\eta^{\mu\nu},\,\, \eta^{\mu\nu}=\diag(1,-1,-1,-1)=\eta_{\mu\nu}.
\EEQ
The $\gamma^{\,5}$ matrix has the properties
\BEQ \hspace{-5mm}
 \gamma^{\, 5}=\ci\gamma^{\,0}\gamma^1\gamma^{\,2}\gamma^{\,3} ,
 \qquad  \{\gamma^{\,\mu}, \gamma^{\, 5}\}=0,
 \qquad (\gamma^{\,5})^2=1.
\EEQ
Left and right handed chiral projectors are, respectively,
\BEQ 
P_L=\half(1- \gamma^{\, 5}),\qquad P_R=\half(1+ \gamma^{\, 5}) . \qquad 
\EEQ
Chiral left handed fields are  $\nu_{L}=P_L\nu$ and chiral right handed ones  $\nu_{R}=P_R\nu$. The projections are
 orthogonal, viz. $P_RP_L=P_LP_R=0$, while $P_L^2=P_L$ and $P_R^2=P_R$.

Hermitian conjugation brings $\gamma^{\,0}{}^\dagger=\gamma^{\,0}$, $\gamma^{\,i}{}^\dagger=-\gamma^{\,i}$,  summarized as
\BEQ
\gamma^{\,\mu}{}^\dagger=\gamma^{\,0}\gamma^{\,\mu}\gamma^{\,0}=\gamma_\mu =\eta_{\mu\nu}\gamma^\nu, \quad 
\gamma^{\,\,5\dagger}= \gamma^{\, 5}.
\EEQ

The charge conjugation matrix $\cC$ has the properties
\BEQ
 \cC^\dagger=\cC^{-1} , \quad \cC^T=-\cC .
 \EEQ
 It is defined up to an overall phase factor, which plays no physical role, and connected to transpositions, 
 \BEQ
\gamma^{\,\mu}{}^T=-\cCdag\gamma^{\,\mu} \cC ,\quad
 \gamma^{\, 5}{}^T=\cCdag \gamma^{\, 5} \cC .
\EEQ

The Pauli matrices are
\BEQ
\sigma_0=\begin{pmatrix}1 &0 \\ 0& 1\end{pmatrix},\quad
\sigma_1=\begin{pmatrix}0 &1 \\ 1& 0\end{pmatrix},\quad
\sigma_2=\begin{pmatrix}0 &-i \\ i& 0\end{pmatrix},\quad
\sigma_3=\begin{pmatrix}1 & 0 \\ 0& -1\end{pmatrix}.
\EEQ

The charge conjugate of any spinor $\nu$ has the properties
\BEQ
& \nu^\cc=\cC \bar\nu^T=-\gamma^{\,0}\cC\nu^{\dagger T}, \quad 
& \nu^{\cc\dagger}=-\nu^T\cC^\dagger\gamma^{\,0},\quad \qquad 
\nu=-\gamma^{\,0}\cC\nu^{\cc T\dagger},\qquad\qquad 
\nn \\
&\nu^T=-\nu^{\cc\dagger}\gamma^{\,0}\cC, \qquad
& \nu^\dagger=-\nu^{\cc T}\cC^\dagger\gamma^{\,0},\qquad \quad\nu^{T\dagger}=-\cC^\dagger \gamma^{\,0}\nu^\cc.
\EEQ
For four-component spinors $\nu_i$ and $\nu_j$ (with $i=j$ allowed) the contraction
 $\nu_i^T\cC^\dagger\nu_j=\sum_{k,l=1}^4\nu_i{}_k\cC^\dagger_{kl}\nu_j{}_l$ is a nonvanishing scalar, 
 since the fermion fields $\nu_{i,j}$ anticommute and $\cC$ is antisymmetric. The relation
\BEQ
(\nu_i^T\cC^\dagger\nu_j)^\dagger=\nu_j^\dagger\cC\nu_i^{T\dagger}=\nu_j^{\cc T}\cC^\dagger\nu_i^\cc ,
\EEQ
assures that the Majorana mass Lagrangian (\ref{LmMR}) is hermitean.

In the chiral representation the $\gamma$ and $\cC$ matrices have the 2$\times2$ blocks
\BEQ && \hspace{-2mm}
\gamma^0=\begin{pmatrix} 0& -\sigma^0\\-\sigma^0&0\end{pmatrix} , \quad
\gamma^i=\begin{pmatrix} 0& \sigma^i\\-\sigma_i&0\end{pmatrix} , \quad
 \cC=\begin{pmatrix} -i\sigma^2& 0\\ 0& i\sigma^2\end{pmatrix} , \nn\\&&
\gamma^5=\begin{pmatrix} \sigma^0& 0\\ 0&-\sigma^0\end{pmatrix}  , \quad
P_L=\begin{pmatrix} 0& 0\\ 0&\sigma^0\end{pmatrix}  , \qquad
P_R=\begin{pmatrix} \sigma^0& 0\\ 0&0\end{pmatrix} .
\EEQ

\section*{Appendix B: The standard model with sterile neutrinos}

For completeness we present the Lagrangian of the standard model with Dirac-Majorana neutrinos  in a compact form 
(leaving out the strong CP violating term),
\BEQ \label{DMnSM}
\hspace{-10mm} \cL &=&
-\frac{1}{4}B^{\mu\nu}B_{\mu \nu} -\frac{1}{4}{\vA}^{\mu\nu} \vA_{\mu \nu} -\frac{1}{4}{\vG}^{\mu\nu} \vG_{\mu \nu} \hfill
+D_\mu\Phi^\dagger D^{\,\mu}\Phi-\mu^2\Phi^\dagger\Phi -\lambda(\Phi^\dagger\Phi )^2
 \nn\\ &&+\,
 \overline{\vQ_L}\ci\slashed{D}\vQ_L+\overline{\vq^U_R}\ci\slashed{D}\vq^U_R +\overline{\vq^D_R}\ci\slashed{D}\vq^D_R 
+\overline{\vL_L}\ci\slashed{D}\vL_L+\overline{\vell_R}\ci\slashed{D}\vell_R   \nn \\
&&-\,\overline{\vQ_L}\Phi \vY^D\vq^D_R\mmin\overline{\vq^D_R} \vY^{D\dagger}\Phi^\dagger\vQ_L
\mmin\overline{\vQ_L}\tilde\Phi \vY^U\vq^U_R\mmin\overline{\vq^U_R}\vY^{U\dagger} \tilde\Phi^\dagger\vQ_L 
\mmin\,\,\overline{\vL_L}\Phi\vY^\ell\vell_R\mmin\overline{\vell_R}\vY^{\ell\dagger}\Phi^\dagger \vL_L 
  \\  && \label{nuKDterms}
+\,\overline{\vnu_R }\ci\slashed{D}\vnu_R -\overline{\vL_L}\tilde\Phi\vY^\nu\vnu_R
 -\overline{\vnu_R }\vY^{\nu\dagger}\tilde\Phi^\dagger \vL_L
  \label{nuMterms}
 +\half \vnu_R ^T\cCdag \vM^{M\dagger}_R  \, \vnu_R  +\half \vnu_R ^{\cc\,T}\,\cCdag \vM^M_R \,\vnu_R^\cc \,     \,  .       
\EEQ
Up to (\ref{DMnSM}) included,  this represents the SM itself: 
the first line contains  the U(1)$_Y$, SU(2)$_L$  and SU(3)$_C$ gauge fields, respectively, and the Higgs kinetic and potential energy.
The second line contains the kinetic terms for three families of quarks, charged leptons and active neutrinos;
the third line lists the quark  couplings to the Higgs field with $3\times 3$ Yukawa matrices $\vY^{U,D}$ and
the charged lepton couplings to the Higgs field with Yukawa matrix $\vY^\ell$.
Eq.  (\ref{nuKDterms}) exhibits the kinetic term for $3$ sterile neutrinos and the Yukawa couplings between the active leptons, 
the Higgs field and  the sterile neutrinos with a $3\times 3$ Yukawa matrix $Y^\nu$.
Eq. (\ref{nuMterms}) also contains the right handed Majorana mass terms of Eq. (\refl{LmMR}). 

The Diracian limit of the main text refers  a special form for $\vM^M_R$.

The quark doublets in Eq. (\ref{DMnSM}) contain the left handed up, down, charm, strange, top and bottom quarks,
\BEQ &&
\vQ_L=\begin{pmatrix} Q_{1L} \\ Q_{2L}\\ Q_{3L} \end{pmatrix},\qquad 
Q_{1L}=\begin{pmatrix} u_L \\ d_L \end{pmatrix},\quad
Q_{2L}=\begin{pmatrix} c_L \\ s_L \end{pmatrix},\quad
Q_{3L}=\begin{pmatrix} t_L \\ b_L \end{pmatrix}.
\EEQ
The lepton doublets contain the left handed electron, muon and tau  (tau lepton, tauon), and their active neutrinos:
the left handed electron, mu and tau neutrino,
\BEQ && \hspace{-5mm}
 \vL_L=\begin{pmatrix} L_{1L} \\ L_{2L}\\ L_{3L} \end{pmatrix},\qquad 
L_{1 L}=\begin{pmatrix} \nu_{e L}      \\   e_L  \end{pmatrix},\quad
L_{2 L}=\begin{pmatrix} \nu_{\mu L} \\ \mu_L \end{pmatrix},\quad
L_{3 L}=\begin{pmatrix} \nu_{\tau L} \\ \tau_L \end{pmatrix}. 
\EEQ
The right handed quark and lepton singlets are grouped as
\BEQ &&
\vq^U_R=\begin{pmatrix} u_R \\ c_R \\ t_R \end{pmatrix},\qquad
\vq^D_R=\begin{pmatrix} d_R \\ s_R \\ b_R \end{pmatrix} ,\qquad  
\bm{ \ell}_R=\begin{pmatrix} e_R \\ \mu_R \\ \tau_R \end{pmatrix},\qquad
\bm{\nu}_R=\begin{pmatrix} \nu_{1R} \\ \nu_{2R} \\ \nu_{3R} \end{pmatrix} . 
\EEQ
The covariant derivatives $D_\mu$ and $\slashed{D}=\gamma^{\,\mu} D_\mu$ in Eq. (\ref{DMnSM}) contain currents from gauge fields, 
except for $\slashed{D}\vnu_{R}=\slashed{\p}\vnu_{ R}$, since $\vnu_R$ is gauge invariant.
Indeed, the weak currents $J_{CC}$  from Eq. (\ref{Jcc=}) and $J_{NC}$ from Eq. (\ref{Jnc=}) arise as parts of $\overline{\vL_L}\ci\slashed{D}\vL_L$. 
The strong currents from $ \overline{\vQ_L}\ci\slashed{D}\vQ_L$ do not involve the neutrino sector.
In the unitary gauge the normal  and conjugated Higgs doublets read, respectively,
\BEQ\hspace{-5mm}
\Phi=\frac{1}{\sqt}\begin{pmatrix} 0 \\ v+H \end{pmatrix}, \quad  \tilde\Phi=\ci\sigma_2\,\Phi^{\dagger T}=\frac{1}{\sqt}\begin{pmatrix}  v+H \\ 0 \end{pmatrix},
\EEQ
where $v=\sqrt{-\mu^2/\lambda}$ is the vacuum expectation value and $H$ the dynamical Higgs field.

After spontaneous symmetry breaking the  $3\times3$ Dirac mass matrices for the ${\it Down}$ = $\{d,s,b\}$ and 
${\it Up}$ = $\{u,c,t\}$ quarks are  
$\vM^D_q =v \vY^D /\sqt$ and  $\vM^U_q =v \vY^U /\sqt$; for the charged leptons  $\vM^\ell =v \vY^\ell /\sqt$
and the Dirac mass matrix for the active neutrinos is $\vM^\nu =v \vY^\nu /\sqt$ (it is denoted as $M^D$ in the main text).
From unitary transformations of the fields it follows that the matrices $\vM^D_q $ and $\vM^\ell$ can be taken diagonal with the respective particle 
masses as entries. Next, the diagonalization of $\vM^U_q$ is performed with the CKM mixing matrix 
and of $M^D=M^\nu$ with the PMNS mixing matrix of Eq. (\ref{Udecomp}).

\newpage



 \end{document}